\documentclass{article}
\usepackage[margin=1.5in]{geometry}

\usepackage[utf8]{inputenc}
\newcommand{\cH}{\mathcal{H}}
\newcommand{\eeq}{\end{equation}}
\newcommand{\ea}{\end{array}}

\def\eea{\end{eqnarray}}

\def\<{\langle}
\def\>{\rangle}

\def\bZ{\mathbb{Z}}
\def\bR{\mathbb{R}}

\def\cO{\mathcal{O}}
\usepackage{amsmath}
\usepackage{comment}
\usepackage{amssymb}
\usepackage{amsthm}

\theoremstyle{definition}

\usepackage{color}
\usepackage{tikz-cd}
\usepackage{comment}

\usepackage{bbm}
\numberwithin{equation}{section}

\renewcommand{\arraystretch}{2}

\usepackage[colorlinks=true, linktocpage=true, allcolors=blue]{hyperref}

\usepackage{tikz}

\title{\huge Anomalies and Bosonization}

\author{Ryan Thorngren}

\date{\small {\it Department of Condensed Matter Physics, Weizmann Institute of Science} \\
{\it Department of Mathematics, University of California, Berkeley}
\\
\medskip
\today}

\begin{document}

\maketitle

\begin{abstract}
Recently, general methods of bosonization beyond 1+1 dimensions have been developed. In this article, we review these bosonizations and extend them to systems with boundary. Of particular interest is the case when the bulk theory is a $G$-symmetry protected topological phase and the boundary theory has a $G$ 't Hooft anomaly. We discuss how, when the anomaly is not realizable in a bosonic system, the $G$ symmetry algebra becomes modified in the bosonization of the anomalous theory. This gives us a useful tool for understanding anomalies of fermionic systems, since there is no way to compute a boundary gauge variation of the anomaly polynomial, as one does for anomalies of bosonic systems. We take the chiral anomalies in 1+1D as case studies and comment on our expectations for parity/time reversal anomalies in 2+1D. We also provide a derivation of new constraints in SPT phases with domain defects decorated by $p+ip$ superconductors and Kitaev strings, which is necessary to understand the bosonized symmetry algebras which appear.

\noindent

\end{abstract}

\section{Introduction}

A theory is said to have an 't Hooft anomaly if it has a global symmetry $G$ which cannot be gauged while preserving locality of interactions. Anomalies are quantized, so if we can identify an anomaly at weak coupling, it is guaranteed to hold at all energy scales \cite{tHooft:1979rat}. This makes anomalies useful for constraining phase diagrams of condensed matter systems whose long range limit is strongly interacting. Likewise, in high energy theory, anomalies which appear in the UV theory constrain the theory at all energy scales \cite{2008arXiv0802.0634B,theta}.

Anomalies are characterized by the anomaly in-flow mechanism \cite{CALLAN1985427}: although we cannot gauge the $G$ symmetry, we can often formulate these $D$-spacetime-dimensional theories as gauge-invariant boundary conditions for a $G$ gauge theory in $D+1$ spacetime dimensions (the ``anomaly theory"). In simple situations, the anomaly theory has a vanishing coupling and a topological term, written as a density made out of the gauge field $A$:
\begin{equation}\label{e:simpleanom}
    S_{anom}(A) = \int_{D+1} \omega(A).
\end{equation}
In this case, under a gauge transformation $A \mapsto A^g$, $S_{anom}(A)$ begets a boundary variation
\begin{equation}\delta S_{anom}(A) = \int_D \omega_1(A,g).\end{equation}
This variation characterizes how the boundary partition function (the partition function of our theory of interest) coupled to the gauge background $A$ fails to be gauge-invariant. Equivalently, $\omega_1(A,g)$ tells us about a kind of higher projective representation of $G$ on the Hilbert space of our theory \cite{CGW,anomaliesvarious}. In this way, $S_{anom}(A)$ characterizes the anomaly.

Possible anomaly theories $S_{anom}(A)$ have been classified by supposing that $S_{anom}(A)$ is a cobordism invariant of the auxilliary $D+1$-manifold equipped with the gauge field $A$ \cite{2014arXiv1403.1467K}. This classification implies that for all bosonic systems, the anomaly theory can be written in the form \eqref{e:simpleanom}. However, for fermionic systems, it is known that not all anomaly theories $S_{anom}(A)$ can be written as an integral of a local density. For example, this is proven for the 1+1D Kitaev phases in \cite{KTTW}. In this situation, we do not yet have a good understanding of what the anomaly ``means" for the $G$ action on the Hilbert space. The purpose for this paper is to fill in this gap.

The toolbox we use to do so is bosonization. Bosonization/fermionization is a family of correspondences between bosonic and fermionic systems. While bosonization has been properly understood in 1+1D for a long time, bosonization in higher dimensions is new, see \cite{GK,higherbos,CKR}, and we will review it. We will especially make use of the bosonization of $G$-symmetry protected topological (SPT) phases \cite{senthilSPT}, which are nondegenerate, gapped systems with a $G$ symmetry which, when gauged, produces a topological term $S_{anom}(A)$. We will need to extend the known description of these systems in bosonization slightly to include the $p+ip$ superconductors and more general symmetry twists.

At first bosonization seems incompatible with anomaly in-flow. Indeed, it is known that the chiral anomaly (for the $\bZ_2$ chiral fermion parity) of a massless free 1+1D Majorana fermion can only be trivialized (meaning we perturb the system into a gapped, nondegenerate, $\bZ_2$-symmetric ground state) if one takes at least 8 such systems and couples them together \footnote{Only the $\bZ_2$ symmetry is assumed to be preserved, but a coupling exists which breaks the flavor symmetry $SO(8)$ to $SO(7)$ \cite{yoni}.}. In this case we say that the anomaly is order 8. However, the bosonization of the Majorana fermion is the critical Ising chain, which has no anomalous $\bZ_2$ symmetries whatsoever. Further, among 1+1D bosonic systems with a $\bZ_2$ symmetry, all anomalies are only order 2, meaning two copies of any 1+1D bosonic system can be gapped out together while preserving all symmetries and introducing no ground state degeneracy.

The resolution of this puzzle lies in the fact that the bosonization of a fermionic SPT is not typically a bosonic SPT but instead has topological order related to a sum over spin structures. This topological order has certain universal higher magnetic symmetries \cite{highersymm} which come from inserting fermionic probes. When the fermionic anomaly cannot be realized in a bosonic system, this means $G$ is nontrivially extended by this higher gauge symmetry, and so in the bosonization the symmetry algebra is modified in the anomalous theory. An example of this has been seen recently in anomalous 2+1D gapped phases \cite{fidkowskianomalies}. We show a 1-to-1 relationship between the data of this symmetry algebra modification and the data which describes the bulk SPT in terms of fermion decorations, equivalently presenting $S_{anom}(A)$ as a cobordism invariant by the Atiyah-Hirzebruch spectral sequence (AHSS).

As a by-product of our analysis, we derive all the $d_2$ differentials of the AHSS for $\Omega^d(BG,\xi)$ for phases obtained in 3rd bosonization by decoration of $G$-gauge defects by fermionic particles, Kitaev strings, and $p+ip$ membranes. Here $\xi$ is the twisting bundle. We collect them here for convenience:
\begin{equation}d_2^1:H^{d-1}(BG,\Omega^1_{spin}) \to H^{d+1}(BG,U(1)^\xi)\end{equation}
\begin{equation}d_2^1 \nu_1 = \frac{1}{2} Sq^2 \nu_1 + \frac{1}{2} w_2(\xi) \nu_1\end{equation}
\medskip
\begin{equation}d_2^2:H^{d-2}(BG,\Omega^2_{spin}) \to H^d(BG,\Omega^1_{spin})\end{equation}
\begin{equation}d_2^2 \nu_2 = Sq^2 \nu_2 + w_2(\xi) \nu_2 + w_1(\xi) Sq^1 \nu_2\end{equation}
\medskip
\begin{equation}d_2^3:H^{d-3}(BG,\Omega^{3,\xi}_{spin}) \to H^{d-1}(BG,\Omega^2_{spin})\end{equation}
\begin{equation}d_2^3\nu_3 = Sq^2 \nu_3 + w_2(\xi) \nu_3 + w_1(\xi)^2 \nu_3.\end{equation}
The first of these is equivalent to the twisted Gu-Wen-Freed equation recently derived by different means in \cite{LZW} and captures all super-cohomology phases.

I would like to thank Anton Kapustin as well as Itamar Hason and Zohar Komargodski for collaborations on related works and sharing their thoughts on the manuscript, also to Max Metlitski and Robert Jones for sharing an early draft of their work, and to Dave Aasen, Yuji Tachikawa, Ruben Verresen, and Dominic Williamson for enlightening conversations. This work is supported by an NSF GRFP fellowship and a Zuckerman STEM Leadership Fellowship.

\section{Overview of Bosonization}

\subsection{0th and 1st Bosonization}\label{sec1stbos}

Bosonization is a general correspondence between bosonic systems (whose fundamental degrees of freedom are bosonic) and fermionic systems (whose fundamental degrees of freedom are fermionic). We will speak in terms of Lorentz-invariant quantum field theory. In this case a fermionic system just means one whose correlation functions on a spacetime $X$ depend on a choice of spin structure $\eta$ on $TX$ \footnote{More generally in the case of charged fermions, the spin structure is ``twisted" in the presence of a background gauge field, meaning that we have a spin structure on $TX \oplus A^*\xi$, where $A^*\xi$ is a bundle associated to a background gauge field $A$. See Section \ref{secbosspt} below. This captures Spin$^c$ systems, etc.}, while a bosonic theory is one whose correlation functions only depend on a choice of orientation of $X$. Note that in this restricted definition, there are field theories which are neither bosonic nor fermionic, such as those with framing anomalies \cite{witten1989} and those which depend on a $w_3$-structure\footnote{It is possible to define a $w_3$ structure on the lattice using the techniques of Chapter 2 of \cite{thesis} but I do not know a Hilbert-space level motivation for them akin to what \cite{GK} do for spin structures. On the other hand, in field theory they have appeared in \cite{thorngren2015framed}.} \cite{higherbos}. Further, it is possible that a lattice system of fermions gives rise to a bosonic continuum limit. Note also that bosonic systems with fermionic quasiparticles do not need a coupling to spin structure, since the quasiparticles are not created by local operators. For a discussion of the connection between the microscopic Hilbert space and spin structures, see \cite{GK}. We refer to a manifold $X$ admitting a spin structure as a spin manifold whereas a manifold $X$ together with a choice of spin structure $\eta$ will be a spun manifold $(X,\eta)$.

The simplest form of bosonization, which might be called 0th bosonization, is a transformation whereby, given a fermionic partition function $Z_f(X,\eta)$ (perhaps depending on sources or other background fields which we supress in the notation of this section), we form the partition function
\begin{equation}Z_b(X) := \frac{1}{\sqrt{|H_1(X,\bZ_2)|}}\sum_{\eta} Z_f(X,\eta)\end{equation}
by summing over all spin structures $\eta$ of $X$. The normalization is arbitrary, but we choose it because it reproduces common conventions in 1+1D. Because spin structures are locally indistinguishable, $Z_b$ will satisfy cluster decomposition for point operators if $Z_f$ does. This means that for two local operators $\cO(x)$ and $\cO'(y)$ supported near points $x$ and $y$ in $\bR^D$, respectively,
\begin{equation}\lim_{|x-y| \to \infty} \langle \cO(x) \cO'(y)\rangle = \langle \cO(x)\rangle\langle \cO'(y)\rangle.\end{equation}
Summing over the spin structures on each side does not change this equation because there is only spin structure on $\bR^D$. Thus, $Z_b$ is a local bosonic field theory. Further, unitarity (reflection positivity) is also unaffected by this transformation.

One can think of 0th bosonization, ie. summing over the spin structures, as (dynamically) gauging fermion parity. This transformation works in any dimension but typically loses some global information about the fermionic theory.

In particular, $Z_b$ will typically have some topological degeneracy since the number of spin structures on $X$ is $|H^1(X,\bZ_2)|$. For example, the bosonization of the trivial fermionic theory, for which $Z(X,\eta) = 1$ for all spun manifolds $(X,\eta)$ is (non-canonically) equivalent to the $\bZ_2$ gauge theory on spin manifolds. Note that the bosonization defines $Z_b(X)$ only for spin manifolds $X$, so the global structure of the 0th bosonization is inherently ambiguous on non-spin manifolds.

An innovation of \cite{GK} is to make bosonization invertible by including neutral probe fermions. One can think of these as Wilson loops $W(\eta,\gamma)$ for the spin structure $\eta$, where $\gamma \in Z_1(X,\bZ_2)$ is the worldline. Accordingly we define the 1st bosonization
\begin{equation}\label{e:1stbos}
    Z_b(X,\gamma) := \frac{1}{\sqrt{|H_1(X,\bZ_2)|}}\sum_{\eta} Z_f(X,\eta) W(\eta,\gamma),
\end{equation}
where, since $W(\eta,\gamma)$ depends only on the spin structure, we may pull it out from the integral over all other fields which computes $Z_f$ \footnote{This bosonization continues to satisfy cluster decomposition because we have just made extended operator insertions to the 0th bosonization.}. We will discuss $W(\eta,\gamma)$ in more detail later, but for now let us note a key equation this Wilson line must satisfy is
\begin{equation}\label{e:torsor}
    W(\eta + \lambda,\gamma) = W(\eta,\gamma) (-1)^{\int_\gamma \lambda},
\end{equation}
where $\lambda \in H^1(X,\bZ_2)$ (this equation implies $W(\eta,\gamma)$ determines $\eta$). From this it follows
\begin{equation}\label{e:torsor2}
W(\eta,\gamma) W(\eta',\gamma)^{-1} = (-1)^{\int_\gamma \eta - \eta'}\end{equation}
depends only on the homology class of $\gamma$ and further
\begin{equation}\label{e:torsor3}
\sum_{[\gamma] \in H_1(X,\bZ_2)} W(\eta,\gamma) W(\eta',\gamma)^{-1} = |H_1(X,\bZ_2)|\ \delta(\eta - \eta'),
\end{equation}
Thus, we can invert \eqref{e:1stbos} by summing over homology classes of probe fermion insertions:
\begin{equation}\label{e:1stfer}
Z_f(X,\eta) = \frac{1}{\sqrt{|H_1(X,\bZ_2)|}}\sum_{[\gamma]} Z_b(X,\gamma) W(\eta,\gamma)^{-1}.
\end{equation}
This backwards transformation is known as 1st fermionization.

In \cite{GK}, the authors interpreted the worldlines as the domain defects of a $D-2$-form $\bZ_2$ symmetry \cite{highersymm,GKSW}, where $D$ is the spacetime dimension. Indeed for a closed spacetime $X$, Poincar\'e duality allows us to express the homology class of the probe worldlines as a cocycle $C \in Z^{D-1}(X,\bZ_2)$, which can be thought of as a gauge background for a $D-2$-form $\bZ_2$ symmetry. Gauge invariance corresponds to homotopy invariance of the probe worldlines, which typically fails, see below. In any case we re-write \eqref{e:1stbos} as
\begin{equation}\label{e:1stbos2}
Z_b(X,C) := \frac{1}{\sqrt{|H_1(X,\bZ_2)|}}\sum_{\eta} Z_f(X,\eta) W_D(\eta,C),
\end{equation}
where we introduce notation which makes the dependence of $W_D(\eta,C)$ on the spacetime dimension especially apparent, since $W_D(\eta,C)$ may only be evaluated on a $D-2$ form $C$.

Thus, the inverse transformation \eqref{e:1stfer} is analogous to how a $\bZ_2$ gauge theory may be ``ungauged", that is--reduced to a theory with global $\bZ_2$ symmetry, by gauging the magnetic $D-2$-form $\bZ_2$ symmetry whose charges are the holonomies of the $\bZ_2$ gauge theory \cite{highersymm,bhardwaj2018finite}. In that case, the ``kernel" of the transformation, which replaces $W(\eta,C)$ in \eqref{e:1stbos} and \eqref{e:1stfer}, is $(-1)^{\int A \cup C}$. We will return to this in 1+1D especially in Section \ref{s:kitaev}.

The authors of \cite{GK} also showed, by studying the gauge transformation properties of $W_D(\eta,C)$, equivalently of $W(\eta,\gamma)$ when $\gamma$ is continuously varied, that the $D-2$-form symmetry of the bosonization is anomalous when $D > 2$, with anomaly
\begin{equation}\label{e:1stbosanom}
    \frac{1}{2}Sq^2 C \in H^{D+1}(B^{D-1}\bZ_2,U(1)).
\end{equation}
This anomaly is nontrivial on general manifolds, but is trivializable on spin manifolds and can be trivialized given a choice of spin structure. In fact, one can think of $W_D(\eta,C)$ as an explicit trivialization. This calls to mind a picture of the fermionization \eqref{e:1stfer} as a slab, with the dynamical boson degrees of freedom and $Z_b$ on one side and the spin structure and $W_D$ on the other, which relates this gauge picture of bosonization to the ``back wall" construction \cite{supertube}. Any bosonic theory with a $D-2$-form $\bZ_2$ symmetry with the above anomaly may be fermionized. We discuss this anomaly in more detail in Section \ref{s:trivphase}.

In \cite{CKR,CK}, the authors also showed how to make the 1st bosonization transformations explicit in general lattice systems. We expect likewise that the results presented here can also be expressed on the lattice, given adequate combinatorial finesse.

In Section \ref{sechigherbos} we will also discuss higher bosonizations, which are defined by inserting higher-dimensional fermionic probes such as Kitaev wires and $p+ip$ superconductors before summing over spin structures. In all cases we are just ``gauging fermion parity" but while identifying (higher) ``magnetic" symmetries of the resulting ``gauge" theory by inserting probes.

\subsection{Some Gapped 1+1D Examples and Convolution}\label{s:kitaev}

In this section we discuss 1st bosonization of the fixed point models of Fidkowski-Kitaev \cite{FidkowskiKitaev}, classified by $\nu \in \bZ_8$, which have particularly simple partition functions. To see all 8 of the distinct phases in bosonization, we will need to generalize \eqref{e:1stbos} to work for non-orientable (hence non-spin) manifolds by modifying the sum to be over Pin$^-$ structures, which are the proper generalization of spin structure to fermions with a time reversal symmetry $T^2 = 1$, and by defining $W(\eta,C)$ for these structures to have the right properties. We discuss these pin structures and other ``twisted spin structures" in Section \ref{secGbos} below.

The goal is to explain the relationship between our bosonization transformation \eqref{e:1stbos} and more standard bosonizations of these phases. We will observe that our 1st bosonization is Kramers-Wannier dual to the usual one. This duality is special to 1+1D bosonization, since in higher dimensions the bosonic symmetries are anomalous and we cannot gauge them without introducing a spin structure and hence refermionizing. See \eqref{e:1stbosanom}, which is trivial for $D = 2$. One advantage of our bosonization in 1+1D is that one can construct a convolution product which reveals the $\bZ_8$ group hidden in the bosonic phases.

We will use \eqref{e:1stbos} where $W(\eta,C) = Q_\eta(C)^{-1}$ is the (inverse of the) $\bZ_4$-valued quadratic form of \cite{KirbyTaylor}, section 3, associated to a Pin$^-$ structure $\eta$ on a closed surface. See also \cite{turzillo2018diagrammatic}. We will comment later on the automorphism $Q_\eta(C) \mapsto Q_\eta(C)^{-1}$, which corresponds to $\nu \mapsto -\nu$ in the Fidkowski-Kitaev classification \cite{FidkowskiKitaev}.

All of our calculations rely only on the following three properties of the spin factor:
\begin{equation}Q_{\eta+\lambda}(C) = Q_{\eta}(C) (-1)^{\int \lambda C}.\end{equation}
\begin{equation}Q_\eta(C + C') = Q_\eta(C) Q_\eta(C') (-1)^{\int CC'}\end{equation}
\begin{equation}\label{e:imaginaryfactor}Q_\eta(C)^{-1} = (-1)^{\int w_1 C} Q_\eta(C),\end{equation}
where all products of cocycles are the cup product, eg. $\lambda C := \lambda \cup C$ \cite{Hatcher}, and $w_1 \in H^1(X,\bZ_2)$ is the orientation class of $X$ \cite{milnor1974characteristic}, whose integrals over closed curves measure if that curve is orientable ($\int w_1 = 0$) or not ($\int w_1 = 1$). Thus the third equation says that if the worldline of our fermionic probe, which recall is Poincar\'e dual to $C$, wraps an orientable cycle in $X$, then $Q_\eta(C)$ is a sign, otherwise it is $\pm i$, which are easily checked from the definitions in \cite{KirbyTaylor} and \cite{turzillo2018diagrammatic}. Later, in Section \ref{s:bosanomaly}, we will be interested in explicit torus partition functions and give a table of values for $Q_\eta(C)$, which are all signs.

First let's consider the $\nu = 0$ trivial phase. Its partition function on any closed surface with any spin structure is 1. Using \eqref{e:1stbos}, we find (supressing normalization)
\begin{equation}\label{e:trivbos2d}
Z_b^0(C) = \delta(C) := \begin{cases}
      C = 0 & 1 \\
      C \neq 0 & 0
   \end{cases},\end{equation}
which corresponds to a spontaneous symmetry breaking (SSB) phase. Indeed, the twisted partition functions, ie. those with $C \neq 0$, of an SSB phase suffer an exponentially large penalty in the system size because of the nonvanishing domain wall tension, causing them all to vanish in the infinite volume limit. On the other hand, the infinite volume limit of the untwisted partition function simply counts the number of symmetric ground states, of which there is one.

We can then dualize by gauging $C$:
\begin{equation}\label{e:KW}
    Z(A)' := \sum_{C} Z(C) (-1)^{\int AC}.
\end{equation}
From this we obtain
\begin{equation}Z_b^0(A)' = 1,\end{equation}
a trivial phase. Let us note that this is usual claimed bosonization of this phase. Indeed, the difference between $Z_b(C)$ and $Z_b(A)'$ is that in the first theory, fermion parity is gauged and $C$ couples to a dual symmetry, while in the second, $C$ is gauged and something which behaves more like a local fermion parity appears as the magnetic symmetry associated to $C$, a symmetry to which $A$ couples via the kernel $(-1)^{\int A C}$ in \eqref{e:KW}.

Next up we consider the $\nu = 1$ phase. Its partition function is the Arf-Brown-Kervaire invariant \cite{KTTW}, defined as
\begin{equation}\label{e:ABK}Z_f^1(\eta) = \sum_B Q_\eta(B).\end{equation}
Thus we find
\begin{equation}Z^1_b(C) = \sum_B \sum_\eta Q_\eta(B) Q_\eta(C)^{-1}\end{equation}
\begin{equation}= \sum_B \sum_\eta Q_\eta(B + C) (-1)^{\int BC + w_1 C}\end{equation}
\begin{equation}= \sum_B (-1)^{\int BC + w_1C} \delta(B+C)\end{equation}
\begin{equation}= (-1)^{C^2 + w_1 C}\end{equation}
\begin{equation}= 1\end{equation}
using the Wu formula $C^2 = w_1 C$. It follows $Z_b^1(A)' = \delta(A)$ is the usual SSB phase.

Now we move on to $\nu = 2$, for which \cite{KTTW}
\begin{equation}Z_f^2(\eta) = Q_\eta(w_1).\end{equation}
We find
\begin{equation}Z_b^2(C) = \sum_\eta (-1)^{\int w_1 C} Q_\eta(C) Q_\eta(w_1)\end{equation}
\begin{equation}= \sum_\eta Q_\eta(w_1+C) = \delta(w_1 + C)\end{equation}
and so
\begin{equation}Z_b^2(A)' = \sum_C \delta(w_1 + C) (-1)^{\int AC}\end{equation}
\begin{equation}= (-1)^{\int w_1 A}.\end{equation}
This is a bosonic SPT for symmetry $P \times T$, where $P$ is a unitary symmetry and $T$ is time reversal, where $PT = -TP$ on the boundary, analogous to the behavior of fermion parity for the $\nu = 2$ boundary in \cite{FidkowskiKitaev}.

Continuing the exercise we obtain the following table of bosonization correspondences:
\begin{center}
 \begin{tabular}{||c c c c||}
 \hline
 $\nu$ & $Z_f(\eta)$ & $Z_b(C)$ & $Z_b(A)'$ \\ [0.5ex]
 \hline\hline
 0 & 1 & $\delta(C)$ & 1 \\
 \hline
 1 & ${\rm Arf}(\eta)$ & 1 & $\delta(A)$\\
 \hline
 2 & $Q_\eta(w_1)$ & $\delta(w_1+C)$ & $(-1)^{\int w_1 A}$ \\
 \hline
 3 & ${\rm Arf}(\eta)Q_\eta(w_1)$ & $(-1)^{\int w_1^2 + w_1 C}$ & $\delta(A+w_1) (-1)^{\int w_1^2}$ \\
 \hline
 4 & $(-1)^{\int w_1^2}$ & $\delta(C)(-1)^{\int w_1^2}$ & $(-1)^{\int w_1^2}$ \\ [1ex]
 \hline
 5 & ${\rm Arf}(\eta)(-1)^{\int w_1^2}$ & $(-1)^{\int w_1^2}$ & $\delta(A) (-1)^{\int w_1^2}$ \\ [1ex]
 \hline
 6 & $Q_\eta(w_1)(-1)^{\int w_1^2}$ & $\delta(w_1 + C)(-1)^{\int w_1^2}$ & $(-1)^{\int w_1 A + w_1^2}.$ \\ [1ex]
 \hline
 7 & ${\rm Arf}(\eta)^{-1}$ & $(-1)^{\int w_1 C}$ & $\delta(A+w_1)$ \\ [1ex]
 \hline
\end{tabular}
\end{center}
Let us note two symmetries of the table. The first, $\nu \mapsto \nu + 4$, corresponds to stacking with the bosonic SPT phase $(-1)^{\int w_1^2}$ which hosts a Kramers doublet at its boundary. Being a bosonic SPT phase, stacking with it commutes with all bosonizations (it pulls out of the sum), so each column is multiplied by $(-1)^{\int w_1^2}$ along with $\nu \mapsto \nu + 4$. A second, $\nu \mapsto -\nu$ is equivalent to taking $Q_\eta(C) \mapsto Q_\eta(C)^{-1} = Q_\eta(C) (-1)^{\int w_1 C}$. This means we add a topological term $(-1)^{\int w_1 C}$ to $Z_b$. In the gauged theory $Z_b'$ this corresopnds to a redefinition of time reversal symmetry by the internal symmetry $A \mapsto A + w_1$.

Let us complete this section by commenting that the stacking operation (tensor product) of fermionic theories, does not correspond to stacking of bosonic theories. Instead there is a sort of convolution product, analogous to the convolution product which occurs in the Fourier transform. By inspection, we define
\begin{equation}(Z^1_b \star Z^2_b)(C) = \sum_B Z^1_b(B) Z^2_b(B+C) (-1)^{\int B^2 + BC}.\end{equation}
Indeed, if we plug in
\begin{equation}Z^j_b(C) = \sum_\eta Z^j_f(\eta) Q_\eta^{-1}(C),\end{equation}
we find
\begin{equation}(Z^1_b \star Z^2_b)(C) = \sum_B \sum_\eta \sum_{\eta'} Z_f^1(\eta) Z_f^2(\eta') Q_\eta(B)^{-1} Q_{\eta'}(B+C)^{-1} (-1)^{\int B^2 + BC}\end{equation}
\begin{equation}= \sum_\eta \sum_{\eta'} Z_f^1(\eta) Z_f^2(\eta') Q_{\eta'}(C)^{-1} \sum_B (-1)^{\int (\eta - \eta') B}\end{equation}
\begin{equation}= \sum_\eta \sum_{\eta'} Z_f^1(\eta) Z_f^2(\eta') Q_{\eta'}(C)^{-1} \delta(\eta - \eta')\end{equation}
\begin{equation}= \sum_\eta Z_f^1(\eta) Z_f^2(\eta) Q_\eta(C)^{-1},\end{equation}
which we recognize as the bosonization of the stack of fermionic theories: $Z_f^1(\eta) Z_f^2(\eta).$ It's easy to show this product is commutative and its unit is $\delta(C)$. Furthermore, one can check that the bosonic phases in the table form a $\bZ_8$ group. In particular, the $\nu \mapsto \nu + 1$ transformation has the particularly simple form:
\begin{equation}(Z_b \star 1)(C) = \sum_B Z_b(B) (-1)^{B^2 + BC}.\end{equation}
It is easy to show that there is no such convolution product for $Z_b'$, since 1 cannot be a unit for any kernel.

\subsection{1st Bosonization of the Trivial Fermionic Phase}\label{s:trivphase}

In this section, we discuss the 1st bosonization of the trivial fermionic phase in general dimensions. This will be important later for the construction of fermionic SPTs by fermionization. We will see the anomaly of \eqref{e:1stbosanom} and a WZW-like formula for $W(\eta,C)$ which were both derived in \cite{GK}.

The trivial $D$-dimensional fermionic phase has partition function
\begin{equation}Z_f^0(X,\eta) = 1\end{equation}
for all closed spacetime $D$-manifolds $X$ and spin structures $\eta$. Suppressing normalization, its 1st bosonization by \eqref{e:1stbos} is
\begin{equation}Z_b^0(X,C) = \sum_{\eta} W(\eta,C).\end{equation}
By the torsor equation \eqref{e:torsor}, $Z_b^0(X,C) = 0$ for all $[C] \neq 0 \in H^{D-1}(X,\bZ_2)$. Thus, $Z_b^0$ is a $D-2$-form-symmetry breaking state. We saw this already for $D = 2$ in \eqref{e:trivbos2d}. In that case, we also saw $Z_b^0(X,0) = 1$. However, for $D > 2$, we will show $Z_b^0(X,C)$ is \emph{not gauge invariant}, meaning that the partition function in the untwisted sector, with $C = d\lambda$, depends non-trivially on $\lambda$.

In the trivial sector, another consequence of the torsor equation \eqref{e:torsor} is that $W(\eta,d\lambda)$ doesn't depend on the choice of $\eta$. We can thus write
\begin{equation}Z_b^0(X,d\lambda) = W(\eta,d\lambda).\end{equation}
Recall that $C = d\lambda$ is Poincar\'e dual to the worldline $\gamma$ of a probe fermion. $\lambda$ then is Poincar\'e dual to a surface $\Sigma$ with the fermion on its boundary $\gamma = \partial \Sigma$.

In \cite{GK}, eq. (9), a WZW-like formula was given for $W(\eta,C)$ when $X$ is the boundary of an oriented $D+1$-manifold $Z$ (not necessarily spin), which in our notation is
\begin{equation}\label{e:wzw}
W(\eta,C) = (-1)^{\int_X \eta \cup C + \int_Z Sq^2 C + w_2 \cup C},
\end{equation}
where $Sq^2$ is the 2nd Steenrod square and the spin structure $\eta$ is expressed as a cochain with $d\eta = w_2$ (we explored this perspective on spin structures further in \cite{thesis}). This formula works especially well for the case $C = d\lambda$ since we can take $Z = X \times [0,1]$ and $C$ extends by zero to a cocycle on $Z$ (so the other side of the slab doesn't contribute anything). Further, the spin structure $\eta$ on $X$ extends to $Z$ so the $\eta \cup C$ and $w_2 \cup C$ terms cancel by Stokes' theorem. The formula becomes simply
\begin{equation}W(\eta,d\lambda) = (-1)^{\int_Z Sq^2 d\lambda},\end{equation}
which is independent of $\eta$ as claimed. Because $Sq^2$ is a cohomology operation, there is a first descendant \cite{highersymm,thesis} $Sq^2_1 \lambda \in C^D(B^{D-1}\bZ_2,\bZ_2)$ which satisfies (and is defined up to coboundaries by)
\begin{equation}\label{e:steenroddesc}
d Sq^2_1 \lambda = Sq^2 d\lambda.
\end{equation}
We can thus finally write
\begin{equation}\label{e:1stbostriv}
Z^0_b(X,d\lambda) = (-1)^{\int_X Sq^2_1 \lambda}.
\end{equation}
As claimed, this is not gauge-invariant, since it depends nontrivially on $\lambda$. However, it is gauge-invariant on the boundary of a theory with topological term $Sq^2 C$ by \eqref{e:steenroddesc}. The construction of this descendant and more of its properties is discussed in detail in Appendix B of \cite{higherbos}.

\subsection{Bosonization with Global Symmetries and Twisted Spin Structures}\label{secGbos}

The general 1st bosonization relation \eqref{e:1stbos} is easily modified to include background fields or sources. In particular, if we have a fermionic theory with a global $G$ symmetry, we couple it to a background $G$ gauge field $A$ and its bosonization will remain coupled to the same gauge field (and hence enjoy the same global symmetry):
\begin{equation}\label{e:1stGbos}
    Z_b(X,A,C) = \sum_{\eta} Z_f(X,A,\eta) W(\eta,C).
\end{equation}
This works if the total symmetry $G_f = G \times \bZ_2^F$ splits between the bosonic part $G$ and the fermion parity $\bZ_2^F$ and so long as $G$ doesn't contain spacetime-orientation-reversing elements.

In more general cases however, our spacetimes won't come equipped with a spin structure but rather with a kind of ``twisted spin structure" which we must sum over instead. To describe it we introduce the twisting bundle $\xi$, a real vector bundle over $BG$, which is equivalently specified by a real representation of $G$ which in \cite{KTTW} was interpreted as the $G$ representation on fermion bilinears.

A $\xi$-twisted spin structure $\eta$ on a manifold $X$ equipped with a $G$-bundle is a spin structure on $TX \oplus A^*\xi$, where $A^*\xi$ is the pullback of the twisting bundle. Actually $\eta$ only depends on the first and second Stiefel-Whitney classes of $\xi$ \cite{milnor1974characteristic}, namely
\begin{equation}w_1(\xi) \in H^1(BG,\bZ_2) \simeq {\rm Hom}(G,\bZ_2)\end{equation}
\begin{equation}w_2(\xi) \in H^2(BG,\bZ_2) \simeq {\rm Ext}(G,\bZ_2).\end{equation}
The first is a homomorphism $G \to \bZ_2$ which picks out the spacetime-orientation-reversing elements of $G$ (eg. time reversal) and the second defines the extension
\begin{equation}\label{e:ext}\bZ_2^F \to G_f \to G.\end{equation}
In practice, one can often forget about the bundle $\xi$ and just remember the two classes $w_1(\xi)$ and $w_2(\xi)$.

Some examples include
\begin{itemize}
    \item A Pin$^-$ structure arises in the study of fermions with a time reversal symmetry $T^2 = 1$ and is the same as a spin structure on $TX \oplus A^*\xi$ where $\xi$ is the 1-dimensional real representation of $\bZ_2$ generated by the scalar $-1$. We studied this case in Section \ref{s:kitaev}.
    \item A Pin$^+$ structure arises in the study of fermions with a time reversal symmetry $T^2 = (-1)^F$ and is the same as a spin structure on $TX \oplus A^*\xi$ where $\xi$ is the 3-dimensional real representation of $\bZ_2$ generated by the scalar matrix $-1$.
    \item A spin$^c$ structure is the same thing as a $U(1)$ gauge field $A$ and a spin structure on $TX \oplus A^*\xi$ where $\xi$ is the usual 1-dimensional complex representation of $U(1)$. These arise in many systems of physical interest, such as QED, where there is a spin-charge relation \cite{seiberg2016gapped,cordova2018global}.
    \item A spin-$SU(2)$ structure as was studied in \cite{newSU2anom} is the same as a $G = SO(3)$ gauge field with a $\xi$-twisted spin structure where $\xi$ the 3-dimensional real vector representation.
    \item The previous two are the $n = 2,3$ special cases of $G = SO(n)$ with $\xi$ the adjoint representation. This structure naturally occurs along an oriented codimension-$n$ submanifold $X$ in a spin manifold $Z$, where $A^*\xi$ is its normal bundle so $TZ|_X = TX \oplus A^*\xi$.
    \item More generally, an unoriented codimension-$n$ submanifold $X$ of a spin manifold carries a $\xi$-twisted spin structure with $G = O(n)$ and $\xi$ its adjoint representation so $A^*\xi = NX$.
\end{itemize}

To apply our bosonization \eqref{e:1stGbos} to these twisted cases we need an extension of the Wilson line operator $W(\eta,C)$ to $\xi$-twisted spin structures $\eta$. In Section \ref{s:kitaev} we used such an extension for $D = 2$ which was defined in \cite{KirbyTaylor}. In \cite{bhardwaj}, a formula in the Pin$^+$ case for $D = 3$ was given similar to the WZW formula \eqref{e:wzw} by replacing $w_2$ with $w_2 + w_1^2$. Presumably the same extension works in any dimension. In fact, based on matching with a spectral sequence calculation below, one can guess the WZW formula has the following extension
\begin{equation}\label{e:wzw2}
    W(\eta,C) = (-1)^{\int_X \eta \cup C + \int_Z Sq^2 C + (w_2(TZ) + w_1(TZ)^2) \cup C},
\end{equation}
where $Z$ is a $D+1$-manifold with $\partial Z = X$ to which $A$ and $C$ extend and $\eta$ represents the twisted spin structure by \cite{KTTW,thesis}
\begin{equation}d\eta = w_2(TX) + w_1(TX)^2 + A^*w_2(\xi).\end{equation}
The formula \eqref{e:wzw2} also appears in \cite{LZW}, eq. (24). Note that this formula is not completely general, as one cannot always find such a manifold with extension. In fact, in the Pin$^-$ case $D = 2$ we already know that $W(\eta,C)$ can be an imaginary phase if $w_1(TX) \cup C \neq 0$ cf. \eqref{e:imaginaryfactor} and see \cite{turzillo2018diagrammatic} for a detailed discussion of bosonization in this case.

A general construction of $W(\eta,C)$ is still lacking and is an interesting direction for developing the theory. In this paper, however, we will just assume such a $W(\eta,C)$ exists and extends \eqref{e:wzw2} (as it must on abstract grounds). This allows us to define the bosonization of the trivial $\xi$-twisted $G$-symmetric fermionic theory as we did previously in Section \ref{s:trivphase}. That is, we bosonize the theory $Z_f^0(X,A,\eta) = 1$. We find using \eqref{e:wzw2}
\begin{equation}\label{e:gentriv}
Z_b^0(X,A,d\lambda) = (-1)^{\int_X Sq^2_1 \lambda + A^*w_2(\xi) \cup \lambda}.
\end{equation}
This implies an anomaly (by taking $d$)
\begin{equation}\label{e:genanom}
Sq^2 C + A^*w_2(\xi) \cup C,
\end{equation}
which agrees with the anomaly computed in \cite{LZW}. We will see it is also consistent with a spectral sequence computation in the appendix.

We further assume $W(\eta,C)$ satisfies the three torsor equations \eqref{e:torsor}, \eqref{e:torsor2}, \eqref{e:torsor3} so that one can define a fermionization transformation inverse to \eqref{e:1stGbos}:
\begin{equation}Z_f(A,\eta) = \sum_C Z_b(A,C) W(\eta,C).\end{equation}

\subsection{SPT Phases}\label{secbosspt}

In this section we describe bosonization of fermionic symmetry protected topological (SPT) phases and how it relates to fermionic decorated domain wall constructions and to the Atiyah-Hirzebruch spectral sequence. These topics have been explored in great detail (in pieces) in \cite{GuWen,kitaevIpam,GK,BGK,bhardwaj,higherbos,WangGu,LZW}, but not always from the point of view of bosonization. Our main purpose is to review these constructions in the context of bosonization.

For us, an SPT is a gapped, invertible theory with a global symmetry $G$ (this is the total symmetry group $G_f$ modulo fermion parity). It is known \cite{KTTW,FreedHopkins} that such phases are classified for fermionic systems by the so-called spin cobordism groups. These groups are dual to the usual spin bordism groups by Anderson duality.

The simplest case is when $G$ contains only unitary symmetries and the total symmetry splits as $G_f = G \times \bZ_2^F$. In this case elements of the cobordism group, written
\begin{equation}Z \in \Omega^D_{\rm spin}(BG)\end{equation}
are partition functions for closed $D$-manifolds with a spin structure.

More generally, when $G$ contains anti-unitary symmetries and is nontrivially extended by $G_f$, we have to use $\xi$-twisted spin structures as in Section \ref{secGbos}. For there there are also bordism groups and a correponding $\xi$-twisted spin cobordism
\begin{equation}\label{e:classification}
Z \in \Omega^D_{\rm spin}(BG,\xi)
\end{equation}
which classifies $G$-SPT partition functions in the general case. As in Section \ref{secGbos}, the group $\Omega^D_{\rm spin}(BG,\xi)$ depends on $\xi$ only through its Stiefel-Whitney classes $w_1(\xi), w_2(\xi)$.

There is a mathematical device for computing the group $\Omega^D(BG,\xi)$ in terms of the groups
\begin{equation}H^j(BG,\Omega^k_{spin}(\star)), \qquad j+k = D\end{equation}
called the Atiyah-Hirzebruch spectral sequence (AHSS) \cite{AHSS}, where $\Omega^k_{spin}(\star)$ is the group of $k$-spacetime-dimension fermionic invertible phases with no assumed global symmetry.

A natural interpretation of an element of $H^{D-k}(BG,\Omega^k_{spin}(\star))$ is a decoration of a $k$-dimensional spacetime defect of $G$-domain walls with a $k$-spacetime-dimensional fermionic invertible phase. Thus, the AHSS says roughly that we need only specify all these decorations to specify the SPT phase, in line with the intuition of \cite{DecoratedDomainWalls}. Note that orientation-reversing or anti-unitary elements of $G$ act nontrivially on these coefficient groups.

First of all, $\Omega^{-1}_{\rm spin}(\star) = \bZ$, and we recognize $H^{D+1}(BG,\bZ^\xi)$ as  the usual\footnote{Note this is isomorphic to $H^D(BG,U(1)^\xi)$ for suitable definition of cohomology \cite{wagemann3304cocycle}.} group of bosonic phases \cite{dijkgraafwitten,CGLW}. The notation $\bZ^\xi$ indicates coefficients twisted by (the determinant of) $\xi$ (the same bundle which appeared in \eqref{e:classification}), ie. by $w_1(\xi)$. This means the elements $\nu_{-1} \in H^{D+1}(BG,\bZ^\xi)$ satisfy the modified cocycle equation
\begin{equation}\label{e:twisteddiff}d_\xi \nu_{-1} = d \nu_{-1} - 2w_1(\xi) \cup \omega = 0.\end{equation}
The operator $d_\xi$ is known as the ``twisted differential", see Chapter 1 of \cite{thesis} for a review of ordinary twisted cohomology.

The simplest truly fermionic phases, ie. with nontrivial spin structure dependence, come from $\Omega^1_{spin}(\star) = \bZ_2$. This $\bZ_2$ describes the two fermionic phases for a quantum mechanical particle with a unique ground state. This ground state can either be bosonic or fermionic, in the latter case its partition function is $+1$ on the anti-periodically spun circle but $-1$ on the periodically spun circle, meaning its partition function generates $\Omega^1_{spin}(\bZ_2)$. Meanwhile a $D-1$-cocycle for $BG$ defines a 1-spacetime-dimensional domain defect (eg. an intersection of $D-1$-many domain walls or just a single domain wall in $D = 2$), so $\nu_1 \in H^{D-1}(BG,\Omega^1_{spin}(\star))$ describes which particle-like defects of $G$ domain walls get decorated with fermions.

Recall that in 1st bosonization, symmetry defects for the $D-1$-form gauge field $C$ are probe fermions. Thus, we can attempt to construct a phase given by some $\nu_1 \in H^{D-1}(BG,\Omega^1_{\rm spin}(\star))$ by beginning with the bosonized description of the trivial fermionic theory we described in Section \ref{s:trivphase} and then simply replacing $C$ with $C + \nu_1(A)$. Recall from \eqref{e:torsor} the 1st bosonization of the trivial fermionic theory has vanishing partition function for all $[C] \neq 0$. In 1+1D we saw in Section \ref{s:kitaev} it was
\begin{equation}Z_b^0(C) = \delta(C),\end{equation}
but in general it is slightly more complicated to account for the anomaly \eqref{e:1stbosanom}, see \eqref{e:1stbostriv}. Thus, the partition function we obtain simplifies considerably to
\begin{multline}\label{e:guwenpartfun}
Z_f(\eta,A) = \sum_C W(\eta,C) Z_b^0(C + \nu_1(A)) = W(\eta,\nu_1(A)) \exp \left(-2\pi i \int_X \nu_0(A) \right) ,
\end{multline}
where we have added the counterterm $\nu_0(A)$ to ensure that $Z_f(\eta,A)$ is gauge invariant (such a counterterm will receive a contribution from $Sq^2_1$ in \eqref{e:1stbostriv}). We will return to this counterterm in a moment. Compare eqn. (50) of \cite{GK}.

If we just consider decorations of domain defects by these two classes of objects \footnote{This constitutes all $k\le 1$ and is thus a truncation of $\Omega_{spin}$ as a spectrum \cite{freed2008,FreedHopkins}.}, we get a subgroup of the full group of fermionic SPT phases. This subgroup was first constructed (on the lattice) by Gu and Wen \cite{GuWen} while for Lie groups it was considered in \cite{freed2008}, so we call them the Gu-Wen-Freed phases. The Gu-Wen-Freed phases are important for us because they are constructed by 1st fermionization \eqref{e:guwenpartfun}. For more general fermionic SPT phases we will need a generalization, which we consider in Section \ref{sechigherbos}.

Recall from the discussion around \eqref{e:1stbosanom} that $W(\eta,C)$ is not invariant under shifts $C \mapsto C + d\lambda$ but is gauge invariant on the boundary of the $D+1$-dimensional term
\begin{equation}\frac{1}{2} Sq^2 C.\end{equation}
Likewise, in the untwisted case $\xi = 0$, $W(\eta,\nu_1(A))$ is also not invariant under shifts $A \mapsto A + df$ but is gauge invariant on the boundary of
\begin{equation}\frac{1}{2} Sq^2 \nu_1(A).\end{equation}
Thus, for \eqref{e:guwenpartfun} to be $G$-gauge invariant, the counterterm must satisfy the Gu-Wen-Freed equation
\begin{equation}\label{e:guwen}
    d \nu_0(A) = \frac{1}{2} Sq^2 \nu_1(A),
\end{equation}
which was identified as a consistency condition in both \cite{GuWen} and \cite{freed2008}. We see that in the case $\nu_1 = 0$ we recover the bosonic phases with cocycle $\nu_0 \in H^D(BG,U(1))$, which can be identified with $\nu_{-1} \in H^{D+1}(BG,\bZ)$ by
\begin{equation}\nu_{-1} = \frac{1}{2\pi i} d \log \nu_0,\end{equation}
which satisfies
\begin{equation}d\nu_{-1} = Sq^1 Sq^2 \nu_1(A).\end{equation}
However, when $\nu_1 \neq 0$, we see that the ``bosonic part" $\nu_0$ of the SPT partition function is ``half-quantized" in the sense of \eqref{e:guwen}. The AHSS reproduces the same result by comparing with the so-called $d_2$ differential.

In the twisted case, $\xi \neq 0$, the situation is more complicated, and we have less understanding of how to define $W(\eta,\nu_1(A))$, although see \cite{bhardwaj}. However, by comparing with the AHSS we find the proper generalization of the Gu-Wen-Freed equations (see Appendix \ref{s:gu-wen-freed})
\begin{equation}\label{e:gu-wen-freed-twist}
    d_\xi \nu_{-1} = Sq^1_\xi (Sq^2 + w_2(\xi)) \nu_1,
\end{equation}
\begin{equation}d_\xi \nu_0 = \frac{1}{2} (Sq^2 + w_2(\xi))\nu_1,\end{equation}
where $d_\xi$ is the twisted differential of \eqref{e:twisteddiff} and $Sq^1_\xi$ is its associated Bockstein operation. One can also derive this from \eqref{e:wzw2}, as was done in \cite{LZW}, which gives a nice consistency check for \eqref{e:wzw2}.

Let us consider again the Kitaev-Fidkowski phases we discussed in Section \ref{s:kitaev}. These phases enjoyed a $\bZ_2$ time reversal symmetry $T$ with $T^2 = 1$. For us this corresponds to a twist $w_1(\xi) = A$, $w_2(\xi) = 0$ \cite{KTTW}. Since $A$ is a $\bZ_2$ gauge field, this means $A = w_1(TX)$ is fixed by the topology of spacetime \cite{2014arXiv1403.1467K}. This identification allows us to compare with the first column of the table (where we used the shorthand $w_1 = w_1(TX)$). We see that the bosonic phases are $\nu = 0, 4$ with $\nu_0 = 0$, $\frac{1}{2} A^2$, respectively and $\nu_1(A) = 0$. As discussed there we also have
\begin{equation}W(\eta,C) = Q_\eta(C)^{-1}\end{equation}
so identifying the partition function \eqref{e:guwenpartfun} the table we see that $\nu = 2, 6$ are Gu-Wen-Freed phases with $\nu_1(A) = A$. Note that in 1+1D, $W(\eta,C)$ is gauge invariant, so $\nu_0$ simply satisfies $d\nu_0 = 0$, and we see that $\nu = 2$ and $\nu = 6$ differ by shifting $\nu_0$ by the cocycle $\frac{1}{2} A^2$.

We see that the odd $\nu$ phases do not appear among the Gu-Wen-Freed partition functions. We would like to obtain an expression of the partition function like \eqref{e:guwenpartfun} for these and more general fermionic SPTs. To do this by a decoration prescription, we will need to consider decorations by Kitaev wires, $p+ip$-superconductors, or even higher-dimensional fermionic invertible phases. In terms of bosonization, we will need to consider modifying our relation \eqref{e:1stbos} to include these extended objects as probes. We address this in the next section and return to the odd $\nu$ Kitaev phases in Section \ref{s:morespts}.

We note that there is another spectral sequence which has appeared as a computational tool for the spin cobordism, namely the Adams spectral sequence, which leverages the action of the Steenrod algebra. See for example \cite{guo2018fermionic,guo2018time}. It would be very interesting to give a physical picture of this spectral sequence.

\subsection{Higher Bosonization}\label{sechigherbos}

In this section, we consider bosonization relations where we insert higher dimensional fermionic probes such as Kitaev wires (which we studied in Section \ref{s:kitaev}) and $p+ip$ superconductors \cite{volovik} into the fermionic path integral before summing over spin structures. We refer to these generalized relations as higher bosonization because we are climbing the Postnikov tower of the spin cobordism spectrum.

As we did in Section \ref{sec1stbos}, we take $C_1 \in C^{D-1}(X,\bZ_2)$ to be Poincar\'e dual to the worldlines of probe fermions, $C_2 \in C^{D-2}(X,\bZ_2)$ Poincar\'e dual to the worldsheets of Kiteav string probes, and $C_3 \in C^{D-3}(X,\bZ)$ to the worldvolumes of $p+ip$ membrane probes. We write the contribution of these objects to the path integral as $W_D(\eta,C_1)$ (which we studied above), $W_D(\eta,C_1,C_2)$ (we will discuss this below), and $W_D(\eta,C_1,C_2,C_3)$ (less is known about this one), where $D$ is the spacetime dimension. We will see later that if you include some extended probes, you must also include all lower dimensional extended probes. By definition, these satisfy
\begin{equation}\label{e:dimrels}W_D(\eta,C_1) = W_D(\eta,C_1,0) = W_D(\eta,C_1,0,0)\end{equation}
\begin{equation}W_D(\eta,C_1,C_2) = W_D(\eta,C_1,C_2,0).\end{equation}

We use these to define the 2nd bosonization
\begin{equation}\label{e:2ndbos}
Z_b(C_1,C_2) := \# \sum_\eta Z_f(\eta) W_D(\eta,C_1,C_2),
\end{equation}
and 3rd bosonization
\begin{equation}\label{e:3rdbos}
Z_b(C_1,C_2,C_3) := \# \sum_\eta Z_f(\eta) W_D(\eta,C_1,C_2,C_3),
\end{equation}
where the normalization is a convention. In higher dimensions one can include extended probes of worldvolume dimension up to $n$ and thus define the $n$th bosonization analogously. Note that all of these bosonizations are really the same theory as the 0th bosonization (just summing over spin structure), but with more of its symmetries identified. For  instance, from \eqref{e:dimrels}, we find the 2nd bosonization restricts to the first bosonization when we turn off $C_2$. Since 1st bosonization is an invertible transformation, all bosonizations beyond the 0th are as well.

In general the possible fermionic probes are all the invertible fermionic phases. For instance, the Kitaev strings correspond to
\begin{equation}\Omega^2_{spin}(\star) = \bZ_2\end{equation}
and the $p+ip$ membranes to
\begin{equation}\Omega^3_{spin}(\star) = \bZ.\end{equation}
The next smallest fermionic SRE phase one might consider is all the way up in 6+1D dimensions, corresponding to a 6+1D gravitational Chern-Simons-like (or thermal Hall-like) system. For bosonizing higher dimensional systems these extended probes may be important. We see any bosonization of a fermionic theory thus has a very large higher symmetry algebra with $\Pi_n = \Omega^{D-n}_{spin}(\star)$, which makes them very special among generic bosonic theories.

As in Section \ref{secGbos}, one can consider the higher bosonization relations for $\xi$-twisted spin structures as well, given a suitable definition of the spin factors $W_D(\eta,-)$ for $\eta$ a $\xi$-twisted spin structure. So far, we only have a good understanding of this for general twists in $D = 2$, where the Kitaev string worldsheet completely wraps the spacetime. Then its contribution is simply an extra factor of the Arf-Brown-Kervaire invariant of $\eta$, when it can be defined. We will return to this in Section \ref{secmajanom}. However, such spin factors must exist on general grounds and satisfy certain ``Postnikov constraints" which restrict the possible $C_1,C_2,C_3$ one can evaluate them on and also have an anomaly generalizing \eqref{e:genanom}, all of which may be derived from the AHSS.

Let us discuss in detail the contribution $W_D(\eta,C_1,C_2)$ of probe fermions and Kitaev strings required for 2nd bosonization in the untwisted case $\xi = 0$, following \cite{higherbos}. If the Kitaev string worldsheets inherit a spin structure from that of the ambient spacetime, then each of them contributes a factor of their Arf invariant, and $W_D(\eta,C_1,C_2)$ is the product of all these Arf invariants and $W_D(\eta,C_1)$, the contributions from the probe fermions.

However, there is no way to write a constraint on the cohomology class of $C_2$ that ensures the Kitaev worldsheet inherits a spin structure. We need to consider more general situations. For instance, the Kitaev wire admits an anti-unitary symmetry with $T^2 = 1$ \cite{FidkowskiKitaev}, which allows it to be defined on any closed Pin$^-$ surface, as we studied in Section \ref{s:kitaev}. Unfortunately one cannot enforce by an equation like the Gu-Wen-Freed equation even the weaker constraint that $\Sigma$ inherits a Pin$^-$ structure.

The best we can do is, given a local framing of $X$ (ie. a choice of local coordinate systems), we can define a restriction of $\eta$ to $\Sigma$ which is Pin$^-$ away from some isolated singularities. These are the points such that if we restrict $\eta$ to the boundary of a small disc $D$ in $\Sigma$ containing the point, then we see the periodic spin structure on $\partial D$, equivalently a fermionic $\pi$-flux, indicating that the induced spin structure on $\partial D$ does not extend over $D$.

It is known that if the Kitaev wire is compactified on a circle with periodic spin structure, it has a unique, fermionic ground state. Thus, we can imagine that at each of these isolated points, a neutral fermionic particle is created from the string. Recall these are Poincar\'e dual to $C_1$. If we connect up all the singularities on $\Sigma$ with probe fermions we then have
\begin{equation}\label{e:2ndbosconstraint}
dC_1 = Sq^2 C_2 \mod 2,
\end{equation}
and given this constraint we can define $W_D(\eta,C_1,C_2)$ as follows. We remove a small disc around each singularity in $\Sigma$ and thicken each of the fermionic probe worldlines to a tube with periodic spin structure in the small direction, gluing the ends of these tubes to the boundaries of $\Sigma$ minus the small discs. The result is a closed surface with Pin$^-$ structure and we define $W_D(\eta,C_1,C_2)$ as its Arf-Brown-Kervaire invariant (see Section \ref{s:kitaev} and \eqref{e:ABK}).

Thus, the 2nd bosonization $Z_b(C_1,C_2)$ is defined for every background $(C_1,C_2)$ satsifying $dC_2 = 0$ and the equation \eqref{e:2ndbosconstraint}, which we interpret as a Postnikov class for the $D-2$-group $E_{2,D}$ with
\begin{equation}\Pi_{D-2} = \Omega^1_{spin} = \bZ_2\end{equation}
\begin{equation}\Pi_{D-3} = \Omega^2_{spin} = \bZ_2.\end{equation}
For more on how to interpret equations like \eqref{e:2ndpostconstraint} as a Postnikov class see \cite{highersymm,thorngren2015higher}. In \cite{higherbos}, this constraint was also related to the $S$ matrix of the Ising TQFT, and more was said about the fermion parity of knotted Kitaev strings.

An aside, it is basically a coincidence that \eqref{e:2ndbosconstraint} resembles the Gu-Wen-Freed equation \eqref{e:guwen} so closely. Indeed, \eqref{e:2ndbosconstraint} is really about the relations in the symmetry algebra of the bosonized theory (and is valued in $\bZ_2$), while \eqref{e:guwen} is about its anomalies (and is valued in $U(1)$). However, see \cite{thorngren2015higher,maissam,delcamp2018gauge,fidkowskianomalies} for a discussion of $\bZ_2$-valued anomalies related to Postnikov classes.

Further, when we add a background gauge field with a nontrivial twisting bundle $\xi$, we will find that the two extensions of these equations, \eqref{e:gu-wen-freed-twist} and \eqref{e:2ndpostconstraintwisted}, are slightly different. In general, all differentials (the complete set of constraints which the decoration data must satisfy) in the AHSS are made from $Sq^2$ and $Sq^1$ and sometimes there are not so many options \cite{Thom1954}. See the appendix for more information.

One can ask whether there is a WZW-like formula for $W(\eta,C_1,C_2)$ akin to \eqref{e:wzw}. This would be very useful for computations and might give a satisfying derivation of \eqref{e:2ndbosconstraint} including in the twisted case $\xi \neq 0$. However, this seems unlikely because one can show that the Arf-Brown-Kervaire invariant of a Pin$^-$ surface is not the integral of any local quantity \cite{KirbyTaylor,KTTW}. However, it does have an expression in bosonization in terms of $W(\eta,C_1)$, namely \eqref{e:ABK}, which is tantalizing.

Likewise we expect that the background $(C_1,C_2,C_3)$ of the 3rd bosonization will have Postnikov constraints defining a higher group $E_{3,D}$ whose solutions form the most general background for which $W_D(C_1,C_2,C_3)$ may be defined. As yet, no one has attempted a derivation along what we outlined for \eqref{e:2ndbosconstraint}, and this is an interesting direction for future work.

However, we do know on abstract grounds that the Postnikov constraints of $E_{3,D}$ should appear as differentials in the AHSS. In Appendix \ref{s:3rdpostder}, we derive one of these constraints:
\begin{equation}\label{e:p+ipconstraint}
    dC_2 = Sq^2 C_3 \mod 2.
\end{equation}
The physical meaning of this constraint is that $Sq^2 C_3$ indicates where the spin structure projected to the $p+ip$ worldvolumes has a particle-like singularity \cite{Thom1954,higherbos}. This is equivalent to a vortex in the $p+ip$ order parameter \cite{omri}, which is known to carry a Majorana zero mode \cite{volovik}. Accordingly, this singularity must lie at the boundary of a Kitaev string worldsheet, which is the meaning of \eqref{e:p+ipconstraint}.

Because the Kitaev string worldsheets are no longer closed, the presence of $C_3$ must also complicate \eqref{e:2ndbosconstraint}. Verifying this, and constructing the spin factor, will be very important for understanding anomalies of 3+1D fermionic systems by bosonization. We leave this to future work.

Finally, in the appendix we derive the $\xi$-twisted versions of \eqref{e:2ndbosconstraint} and \eqref{e:p+ipconstraint} using the AHSS and find
\begin{equation}\label{eqn2ndbostwisted}
  dC_1 = Sq^2 C_2 + w_2(\xi) \cup C_2 + w_1(\xi) \cup Sq^1 C_2 \mod 2
\end{equation}
in the case $C_3 = 0$, otherwise
\begin{equation}\label{eqn3rdbostwisted}
  dC_2 = Sq^2 C_3 + w_2(\xi) \cup C_3 + w_1(\xi) \cup w_1(\xi) \cup C_2 \mod 2
\end{equation}
in which case \eqref{eqn2ndbostwisted} becomes modified in some unknown way.

\subsection{SPTs Beyond Gu-Wen-Freed}\label{s:morespts}

In this section we will use the higher bosonization relations to construct fermionic SPT phases beyond the Gu-Wen-Freed phases discussed in the previous section. The strategy will be the same however: we consider the 2nd or 3rd bosonization of the trivial fermionic phase to obtain
\begin{equation}Z_b^0(C_1,C_2) \qquad Z_b^0(C_1,C_2,C_3),\end{equation}
and then we shift $C_j \mapsto C_j + \nu_j(A)$ for some $\nu_j \in C^{D-j}(BG, \Omega^j_{\rm spin})$ (this is schematic, see below for some caveats). As in 1st bosonization, $Z_b^0$ vanishes in all non-twisted sectors, so by a gauge choice this is the same as just setting $C_j = \nu_j(A)$ (compare Section \ref{s:trivphase}). We now refermionize to obtain a fermionic SPT constructed from decorations by probe fermions, Kitaev strings, and $p+ip$ membranes. The resulting partition functions are of the form
\begin{equation}\label{eqnSPTpartfun}
  Z_f(\eta,A) = W_D(\eta,\nu_1(A),\nu_2(A),\nu_3(A)) \exp\left(-2\pi i \int_X \nu_0(A)\right),
\end{equation}
where $\nu_0 \in C^D(BG,U(1))$ is a counterterm which ensures $G$ gauge invariance, as in Section \ref{secbosspt}. Compare \eqref{e:guwenpartfun} which is obtained by setting $\nu_2$ and $\nu_3$ to zero. Also see \cite{putrov2017braiding}, which gives several formulas for partition functions of this form.

We see that in order to evaluate $W_D(\eta,\nu_1,\nu_2)$ or $W_D(\eta,\nu_1,\nu_2,\nu_3)$, the $\nu$'s must satisfy the same Postnikov constraints the $C$'s did. To wit, for phases constructed in 2nd bosonization (no $p+ip$ decorations, ie. $\nu_3 = 0$) the condition is
\begin{equation}d\nu_2 = 0\end{equation}
\begin{equation}\label{e:2ndpostconstraint}
d\nu_1 = Sq^2 \nu_2
\end{equation}
in the untwisted $\xi = 0$ case and
\begin{equation}\label{e:2ndpostconstraintwisted}
    d\nu_1(A) = Sq^2 \nu_2(A) + w_2(A^*\xi) \nu_2 + w_1(A^*\xi) Sq^1 \nu_2.
\end{equation}
in general. Actually we find that we cannot simply shift the $C$'s by the $\nu$'s and have the result be valid inputs for $W_D$. Instead the replacement is
\begin{equation}C_2' = C_2 + \nu_2\end{equation}
\begin{equation}C_1' = C_1 + \nu_1 + \chi(C_2,\nu_2),\end{equation}
where $\chi$ is a universal cross-term which satisfies
\begin{equation}d\chi(C_2,\nu_2(A)) = Sq^2 C_2 + Sq^2 \nu_2(A) - Sq^2 (C_2 + \nu_2(A)) \mod 2.\end{equation}
This ensures $C_1'$ satisfies the Postnikov constraint \eqref{e:2ndbosconstraint}. (Note that the correction terms in \eqref{e:2ndpostconstraintwisted} are linear in $\nu_2$ so this works in the general case as well.) One solution for $\chi$ is
\begin{equation}\chi(C_2,\nu_2(A)) = C_2 \cup_{D-1} \nu_2(A),\end{equation}
using the $\cup_i$ products of Steenrod \cite{cohomologyoperations}. This also applies for the group law on the $\nu$'s and yields the group law for fermionic SPT phases once all of the cross-terms are found, see \cite{BGK}.

As with 1st bosonization, $W_D(\eta,\nu_1(A),\nu_2(A))$ is not $G$-gauge invariant, and we must choose the counterterm $\nu_0(A)$ so that the combined partition function $Z_f(\eta,A)$ is. In \cite{higherbos} the authors found that for $\xi = 0$ the anomaly depends on both $\nu_2$ and $\nu_1$ and was computed as a cohomology element
\begin{equation} \tilde Sq^2(C_1,C_2) \in H^{D+1}(BE_{2,D},U(1)),\end{equation}
which restricts to $Sq^2 C_1$ when $C_2 = 0$, yielding the Gu-Wen-Freed equation. In \cite{WangGu}, a different, presumably equivalent expression was given to $\tilde Sq^2(C_1,C_2)$. In summary, we need a $\nu_0 \in C^D(BG,U(1))$ satisfying the higher Gu-Wen-Freed equation
\begin{equation}\label{e:higherguwen}
d\nu_0 = \frac{1}{2} \tilde Sq^2(C_1,C_2) \mod 1.
\end{equation}
See also \cite{brumfiel2018pontrjagin}.

In the twisted case $\xi \neq 0 $, we expect that there will be a stable cohomology operation $\tilde Sq^2_\xi(\nu_1,\nu_2)$ for $\nu_1,\nu_2$ satisfying \eqref{e:2ndpostconstraint} which defines the Gu-Wen-Freed equation by
\begin{equation}d\nu_0 = \frac{1}{2}\tilde Sq^2_\xi(\nu_1,\nu_2),\end{equation}
generalizing \eqref{e:higherguwen}. Such a cohomology operation can be shown to exist but it has not been explicitly constructed. Let us point out however that even in 2+1D, where $d\nu_1$ and $d\nu_2$ are both zero, so that fermion particle number and Kitaev string flux are separately conserved, there is still a nontrivial contribution of $\nu_2$ to the Gu-Wen-Freed equation. It is (generalizing \cite{GK,BGK})
\begin{equation}\label{e:higherguwen3d}
    d_\xi \nu_0 = \frac{1}{2} (Sq^2 + w_2(\xi)) \nu_1 + \frac{1}{2}(w_2(\xi) + w_1(\xi)^2) Sq^1 \nu_1  \qquad D = 2+1,
\end{equation}
which can be determined using AHSS techniques. We leave the explicit verification of this and the Gu-Wen-Freed equations in higher dimensions to future work.

In $D = 2$ what we have is enough to describe all the phases in the $\bZ_8$ Kitaev phases in Section \ref{s:kitaev}. For these Pin$^-$ phases, the Postnikov constraints are very simple
\begin{equation}dC_1 = 0\end{equation}
\begin{equation}dC_2 = 0\end{equation}
and there is no bosonization anomaly so the Gu-Wen-Freed equation is simply
\begin{equation}d\nu_0 = 0.\end{equation}
We can write
\begin{equation}\label{eqn2dspinfactor}
  W_2(\eta,C_1,C_2) = Q_\eta(C_1) {\rm ABK}(\eta)^{C_2},
\end{equation}
where ${\rm ABK}(\eta)$ is the Arf-Brown-Kervaire invariant of $\eta$ and the exponent means that, as $C_2 \in Z^0(X,\bZ_2)$ is a constant, if $C_2 = 1$ we have this term and otherwise we don't\footnote{This spin factor differs by complex conjugation from the one in Section \ref{s:kitaev} but the fermionization convention \eqref{e:1stfer} does as well. This is fixed by exchanging $W$ with $W^{-1}$ everywhere.}. Thus, the odd $\nu$ phases in the $\bZ_8$ have $C_2 = 1$ by comparing with the tabl in Section \ref{s:kitaev} and \eqref{eqnSPTpartfun}. We see $\nu = 1$ has $\nu_1 = \nu_0 = 0$; $\nu = 3$ has $\nu_1 = A$, $\nu_0 = 0$; $\nu = 5$ has $\nu_1 = 0$, $\nu_0 = A^2/2$; and $\nu = 7$ has $\nu_1 = A$, $\nu_0  = A^2/2$. Further discuss of a $\bZ_8$ in 2+1D can be found in Section \ref{secchiralbos}.

More generally, we can consider also turning on $\nu_3$, allowing decoration of symmetry defects by $p+ip$ superconductors. $\nu_3$ is a bit different from $\nu_1$ and $\nu_2$ because it is integer-valued, while $\nu_1$ and $\nu_2$ are integer mod 2-valued. This means that the complex conjugated phases $p+ip$ and $p-ip$ are distinct, so $\nu_3$ describes a cocycle in twisted cohomology
\begin{equation}\nu_3 \in Z^{D-3}(BG,\bZ^\xi),\end{equation}
meaning it satisfies the twisted cocycle equation
\begin{equation}d_\xi \nu_3 = d\nu_3 - 2w_1(\xi) \cup \nu_3 = 0.\end{equation}
Compare \eqref{e:twisteddiff} and \eqref{e:higherguwen3d}.

In 2+1D, when there are no orientation-reversing or anti-unitary elements,
\begin{equation}\nu_3 \in H^0(BG,\bZ^\xi) = H^0(BG,\bZ) = \bZ\end{equation}
simply indicates how many $p+ip$ superconductors there are layered with a torsion SPT to form the SPT corresponding to $\nu_3,\nu_2,\nu_1,\nu_0$, similar to how in 1+1D, a nonzero $\nu_2 \in H^0(BG,\bZ_2)$ indicates the presence of a Kitaev string wrapping space. Equivalently $\nu_3/2$ is the chiral central charge of the boundary modes. When $G$ does contain orientation-reversing or anti-unitary elements, however,
\begin{equation}H^0(BG,\bZ^\xi) = 0\end{equation}
and there is no possibility for a $p+ip$ superconductor.

In 3+1D, with no orientation-reversing or anti-unitary elements,
\begin{equation}\nu_3 \in H^1(BG,\bZ^\xi) = H^1(BG,\bZ) = {\rm Hom}(G,\bZ),\end{equation}
which is zero unless $G$ contains a subgroup isomorphic to $\bZ$, such as a discrete translation symmetry subgroup. For such phases with a nonzero $\nu_3$, we have a stack of $p+ip$ superconductors along the translation. In fact, writing the smallest translation as $t$, $\nu_3(t) \in \bZ$ tells us the number of $p+ip$ superconductors in each layer.

In 3+1D with orientation-reversing or anti-unitary elements, it is possible to have a nonzero $\nu_3$ even when $G$ is finite. For instance, if $G = \bZ_2$ with the nontrivial action on $\bZ$,
\begin{equation}H^1(B\bZ_2,\bZ^\xi) = \bZ_2.\end{equation}
This indicates that the $\bZ_2$ domain wall carries a $p+ip$ superconductor. This is familiar from the study of the 3+1D topological superconductors with $T^2 = (-1)^F$, where there is the possibility of breaking time reversal symmetry in two opposite ways on an otherwise symmetric boundary, yielding a domain wall carrying a $c = 1/2$ mod 1 chiral mode \cite{MFCV,FCV}. We will discuss this phase more below.

We wish to understand how the presence of $p+ip$ defects changes the Postnikov constraints and the Gu-Wen-Freed equation. In Appendix \ref{s:3rdpostder}, we derive the first Postnikov constraint for these $p+ip$ membrane phases:
\begin{equation}\label{e:3rdpostconstraint}
    d\nu_2 = (Sq^2 + w_2(\xi) + w_1(\xi)^2) \nu_3 \mod 2.
\end{equation}
Physically, the right hand side corresponds to certain curve-like singularities in the $p+ip$ worldvolumes where Kitaev strings are created, in fact these are the Majorana zero modes along the worldlines of $p+ip$ vortices \cite{volovik}, cf. the discussion around \eqref{e:p+ipconstraint}.

The Postnikov constraint \eqref{e:3rdpostconstraint} will also complicate the Postnikov constraint for $d\nu_1$ (conservation of fermion parity) and the Gu-Wen-Freed equation for $d\nu_0$ ($G$-anomaly vanishing). These equations have not yet been worked out and we leave it to future study. However, for $D = 3+1$ and time reversal symmetry with $T^2 = (-1)^F$, we have $Sq^2 \nu_3 = 0$ since $\nu_3$ is only a 1-cocycle (since $p+ip$ defects are codimension 1 they always have $w_2(NV) = 0$) and $w_2(\xi) + w_1(\xi)^2 = 0$. In this case, we simply have
\begin{equation}\label{e:T2=-1post}
    d\nu_2 = 0 \mod 2,
\end{equation}
meaning there are no vortices present on the $p+ip$ defects and so the Kitaev strings are conserved. In this case one can also find that \eqref{e:2ndpostconstraintwisted} is unmodified, although we still do not know the Gu-Wen-Freed equation in this case. However, for understanding the modification of the time reversal symmetry algebra in 3rd bosonization, this will be enough.

% For $D\le 4$, 2+1D volumes $V$ are codimension 0 or 1, and so $w_2(NV) = 0$. It follows that if $w_2(\xi) = 0$, then $\nu_3$ does not have any nontrivial AHSS differentials, so we have in general
% \begin{equation}d_G \nu_3 = 0\end{equation}
% \begin{equation}d\nu_2 = 0 \mod 2\end{equation}
% \begin{equation}d\nu_1 = Sq^2 \nu_2 \mod 2\end{equation}
% \begin{equation}d_G\nu_0 = \frac{1}{2} \tilde Sq^2 (\nu_1,\nu_2,\nu_3)\end{equation}
% where $d_G$ denotes the differential for twisted coefficients. In particular, neither \eqref{e:2ndbosconstraint} nor the Gu-Wen-Freed equation is modified. A similar phenomenon occurs in 2+1D, where $\nu_2$ does not enter the Gu-Wen-Freed equation. However, if we have a twist with $w_2(\xi) \neq 0$, then we can expect
% \begin{equation}d\nu_2 = w_2(\xi) \nu_3 \mod 2,\end{equation}
% as happens in 2+1D, since $w_2(\xi)$ is a bulk fermionic $\pi$-flux independent of the embedding of $V$.

\section{1st Bosonization with Boundary}

We wish to consider the problem of bosonization for systems with boundaries. In this situation, we have a fermionic theory defined on a spun $D$-manifold $(X,\eta)$ with boundary $\partial X = Y$. Along the boundary we have a splitting
\begin{equation}TX = TY \oplus NY,\end{equation}
where $NY$ is a line bundle which is trivialized by identifying a neighborhood of $Y$ with the ``collar" \begin{equation}Y \times [0,1] \subset X,\end{equation}
where $Y \times 0 = \partial X$, and choosing a coordinate system where the first $n-1$ coordinates are parallel to $Y$ and the last coordinate is the perpendicular coordinate along the interval $[0,1]$. With this choice of trivialization of $NY$, a spin structure $\eta$ on $X$ canonically determines a spin structure $\eta|_Y$ on $Y$. Thus we will assume that the fermionic degrees of freedom localized to the boundary couple to the spin structure $\eta|_Y$, so the fermionic theory has partition function $Z_f(X,\eta)$ with no extra choices.

Recall the 1st bosonization is constructed by introducing fermion probe particles, and using Poincar\'e duality to relate the worldlines of these particles to the domain defects of an anomalous higher symmetry. We do the same in the presence of a boundary, except now the proper tool is Poincar\'e-Lefschetz duality, which says that the 1-cycle $\gamma$ is equivalent to a pair of gauge fields:
\begin{equation}C \in Z^{D-1}(X,\bZ_2) \qquad B \in C^{D-2}(Y,\bZ_2)\end{equation}
such that
\begin{equation}dB = C|_Y \mod 2.\end{equation}
This means that $C$ is an $D-1$-form gauge field on $X$ and $B$ is a Dirichlet boundary condition which says $C|_Y$ is gauge-equivalent to the zero connection. Note that a Neumann boundary condition would allow free flow of fermionic worldlines through the boundary, and fermion parity would not be conserved.

The duality lets us define the spin factor by
\begin{equation}W_D(\eta,C,B) = W(\eta,\gamma).\end{equation}
Then we define the 1st bosonization with boundary as
\begin{equation}\label{e:bosonizationboundary}
    Z_b(C,B) = \sum_\eta Z_f(\eta) W_D(\eta,C,B).
\end{equation}
As before, $W_D(\eta,C,B)W_D(\eta',C,B)^{-1}$ depends only on the (relative) cohomology class $[C,B] \in H^{D-1}(X,Y,\bZ_2)$ and we have
\begin{equation}\sum_{[C,B]} W_D(\eta,C,B)W_D(\eta',C,B)^{-1} = \# \delta(\eta - \eta').\end{equation}

Further, $W_D(\eta,C,B)$ connects $W_D(\eta,C)$ and $W_{D-1}(\eta|_Y,B)$ as follows. If $\gamma$ is supported away from the boundary, then $C|Y = 0$, and $B = 0$ and so
\begin{equation}W_D(\eta,C,0) = W_D(\eta,C)\end{equation}
Oppositely, when $\gamma$ is supported only on the boundary, then $C = 0$, $dB = 0$ and we have
\begin{equation}\label{e:descent}
W_D(\eta,0,B) = W_{D-1}(\eta|_Y,B).
\end{equation}
This means that the bosonization of the bulk-boundary system is closely related to the bosonization of the boundary theory, but note that even when the bulk is the trivial theory, the bosonization will only sum over spin structures on $Y$ which extend to $X$. In particular, the bulk-boundary bosonization is unaware of any fermionic SRE phase on the boundary. This will become an important point for us later. We mention that another consistent bosonization of $Y$ when $X$ carries the trivial theory would be to sum over all bounding spin structures (``summing" over the filling $X$). We won't use this bosonization in this paper though.

\subsection{1st bosonization of Gu-Wen-Freed anomalies}

Now we consider 1st bosonization for a Gu-Wen-Freed $G$-SPT with symmetric boundary. This means we have a $G$-SPT on $X$ described by $\nu_1, \nu_0$ satisfying the Gu-Wen-Freed equation \eqref{e:guwen}, and a $G$-symmetric boundary theory on $Y = \partial X$, which is typically gapless. We wish to bosonize the whole thing.

Let $A$ denote the background $G$ gauge field $A \in Z^1(X,G)$, with free boundary conditions on $Y$ (which we can do as long as the $G$ symmetry is preserved in the boundary theory). Recall $\nu_1(A)$ represents a particle-like defect binding a neutral fermion. The condition that these worldlines carry a $\bZ_2$ quantum number (meaning they have no free ends) translates in Poincar\'e-Lefschetz duality to
\begin{equation}
d(C + \nu_1(A)) = 0 \mod 2
\end{equation}
\begin{equation}(C + \nu_1(A))|_Y = dB \mod 2.\end{equation}

The first follows from $dC = d\nu_1(A) = 0$ mod 2, but the second is a key equation. If we consider the case $C = 0$, meaning the probe fermions are restricted to the boundary, then it becomes
\begin{equation}dB = \nu_1(A) \mod 2.\end{equation}
This implies when $[\nu_1(A)] \neq 0 \in H^{D-1}(BG,\bZ_2)$ that the global $G$ symmetry is nontrivially extended by the $B^{D-3}\bZ_2$ symmetry which couples to $B$!

For example, when $D = 3$, we are studying anomalies of fermionic systems in 1+1D, the usual setting for bosonization. We have found that if the $G$ anomaly of the fermionic system is of Gu-Wen-Freed type, but not equivalent to any bosonic $G$ anomaly, it means that when we bosonize, obtaining a 1+1D bosonic theory with a global $\bZ_2$ symmetry, that $G$ is nontrivially extended by this $\bZ_2$ symmetry. This resolves the apparent contradiction that fermionic systems have more $G$ anomalies than bosonic systems but are supposed to be equivalent by bosonization/fermionization.

\subsection{1+1D Chiral $U(1)$ Anomaly}\label{s:bosanomaly}

Before we get too lost, let's discuss a concrete example of this.

Consider a free massless Dirac fermion in 1+1D. The fermion number of this theory is a conserved integer $N$, which may be considered a sum of occupation numbers $N = N_L + N_R$ from the left-moving and right-moving sectors, respectively. It's known that the corresponding chiral $U(1)_{L,R}$ symmetries can only be consistently coupled to a background gauge field on the boundary of a $U(1)$ SPT phase with Chern-Simons level $\pm 1$. We wish to describe this situation in bosonization. Here $D = 3$.

First of all, the level 1 Chern-Simons term contains a hidden dependence on a spin structure \cite{BelovMoore}, which in bosonization is encoded in a nontrivial $\nu_1$. To see this, we study a $2\pi$-flux for the $U(1)$. An odd level Chern-Simons term decorates the bare $2\pi$-flux with an odd electric charge, so the physical $2\pi$-flux is a fermion. Thus, the proper spin factor to use in bosonization is
\begin{equation}W_3(\eta,C + c_1(A),B),\end{equation}
where $c_1(A)$ is the first Chern class of $A$. In other words, $\nu_1(A) = c_1(A)$ mod 2. This means that in the boundary theory, we expect the $U(1)$ group relation will only hold up to the $\bZ_2$ gauge symmetry of $B$. Denoting the boundary $U(1)$ charge $Q$ and the $\bZ_2$ charge $s$, $\nu(A) = c_1(A)$ mod 2 means
\begin{equation}e^{2\pi i Q} = (-1)^s.\end{equation}

We will verify this prediction by studying the torus partition functions of the Dirac fermion. On a torus, the spin factor $W_2(\eta,B)$ for $B$ one of the four $\bZ_2$ gauge backgrounds $+/+, +/-, -/+, -/-$\footnote{Our notational convention specifies the gauge or spin holonomy around the space/time cycles, respectively.} and for $\eta$ one of the four spin structures $AP/AP, AP/P, P/AP, P/P$ is encoding in the follow matrix:

  \begin{equation}
\renewcommand\arraystretch{1.3}
\left[
\begin{array}{c|cccc}
  W(\eta,C) & AP/AP & AP/P & P/AP & P/P \\
  \hline
  +/+ & 1 & 1 & 1 & 1 \\
  +/- & 1 & 1 & -1 & -1 \\
  -/+ & 1 & -1 & 1 & -1 \\
  -/- & -1 & 1 & 1 & -1
\end{array}
\right]
\end{equation}

The four partition functions of the free Dirac fermion on a torus with shape parameter $\tau$, and $q = e^{2\pi i \tau}$ are \cite{francesco2012conformal}
\begin{equation}Z_f(AP/AP) = |\chi_{1,1}(q) + \chi_{2,1}(q)|^4 = \frac{1}{|\eta(\tau)|^2} \sum_{a,b \in \bZ} q^{\frac{1}{2}a^2} \bar q^{\frac{1}{2}b^2} = \bigg| \frac{\theta_3(\tau)}{\eta(\tau)} \bigg|^2\end{equation}
\begin{equation}Z_f(AP/P) = 4 |\chi_{1,2}(q)|^4 = \frac{1}{|\eta(\tau)|^2} \sum_{a,b \in \bZ} (-1)^{a+b} q^{\frac{1}{2}a^2} \bar q^{\frac{1}{2}b^2} = \bigg| \frac{\theta_4(\tau)}{\eta(\tau)} \bigg|^2\end{equation}
\begin{equation}Z_f(P/AP) = |\chi_{1,1}(q) - \chi_{2,1}(q)|^4 = \frac{1}{|\eta(\tau)|^2} \sum_{r,s \in \bZ + \frac{1}{2}} q^{\frac{1}{2}r^2} \bar q^{\frac{1}{2}s^2} = \bigg| \frac{\theta_2(\tau)}{\eta(\tau)} \bigg|^2\end{equation}
\begin{equation}Z_f(P/P) = \frac{1}{|\eta(\tau)|^2} \sum_{r,s \in \bZ + \frac{1}{2}} (-1)^{r-s} q^{\frac{1}{2}r^2} \bar q^{\frac{1}{2}s^2} = 0.\end{equation}

Using the table above, we find, for example the untwisted partition function of the bosonization:
\begin{equation}Z_b(+/+) = \frac{1}{|\eta(\tau)|^2} \sum_{m,n \in \bZ} q^{\frac{1}{2}(n/2 + m )^2} \bar q^{\frac{1}{2}(n/2 - m)^2}.\end{equation}
This is the partition function of the compact boson at radius $R = 2$, as is well-known to be expected \cite{francesco2012conformal}. To figure out how the background gauge field should couple to the compact boson, we can use the bosonization relations to compute
\begin{equation}Z_b(+/-) = \frac{1}{|\eta(\tau)|^2} \sum_{m,n \in \bZ} (-1)^n q^{\frac{1}{2}(n/2 + m )^2} \bar q^{\frac{1}{2}(n/2 - m)^2}.\end{equation}
Thus, in the eigenbasis $|n,m\rangle$ described by this partition function, the $\bZ_2$ charge operator is diagonal, with $s = n$.

To identify how the chiral $U(1)_L$ symmetry acts on the compact boson, we perform the bosonization transformation in the presence of a flat gauge background $1/\theta$ for $A$, that is with a twist $e^{i\theta Q}$ along the temporal cycle and untwisted in the spatial direction. For the fermion, this means we add a phase
\begin{equation}e^{i \theta a}\end{equation}
in the sum over characters computing the $AP/\star$ partition functions, while for the $P/\star$ partition functions we use
\begin{equation}e^{i \theta r}.\end{equation}
Note that in this later expression, since $r$ is a half-integer, there is a choice of branch of the logarithm. We choose $\theta \in [0,2\pi)$, but this choice does not affect our computation of the anomaly.

Taking these partition functions through the bosonization transformations, we find
\begin{equation}Z_b(+/+;0/\theta) = \frac{1}{|\eta(\tau)|^2} \sum_{m,n \in \bZ} e^{i \theta (n/2+m)}q^{\frac{1}{2}(n/2 + m )^2} \bar q^{\frac{1}{2}(n/2 - m)^2}.\end{equation}
Thus in the $|n,m\rangle$ basis, we can identify the chiral $U(1)_L$ charge with
\begin{equation}Q = n/2 + m.\end{equation}
In particular we find
\begin{equation}e^{2\pi i Q} = (-1)^s\end{equation}
as expected: $U(1)_L$ is extended by $\bZ_2$ in bosonization.

Given $\nu_1(A) = c_1(A)$ mod 2, there are still infinitely many choices for $\nu_{-1}(dA/2\pi)$, equivalently $\nu_0(A)$, satisfying the Gu-Wen-Freed equation:
\begin{equation}d\nu_0(A) = \frac{1}{2} c_1(A)^2 \mod 1,\end{equation}
given by
\begin{equation}\nu_0(A) = \frac{k}{4\pi} A \frac{dA}{2\pi} \qquad k \in 2\bZ + 1,\end{equation}
which we recognize as the odd-level $U(1)$ Chern-Simons terms.\footnote{These are schematic expressions which must be treated carefully, eg. using differential cocycles, to account for torsion situations where $c_1(A)$ is nontrivial but $dA = 0$.}

To determine the level, we need to study the charge of the magnetic flux. We can obtain the partition function
\begin{equation}Z_b(+/+,\theta/\theta) = \frac{1}{|\eta(\tau)|^2} \sum_{m,n \in \bZ} e^{ i \theta (n/2 + m + \theta/4\pi)} q^{\frac{1}{2} (n/2 + m + \theta/2\pi)^2} \bar q^{\frac{1}{2} (n/2 - m)^2}\end{equation}
from a modular transformation of $Z_b(+/+,0/\theta)$. This partition function indicates that the $\theta$ flux operator carries $\theta/4\pi$ charge, indicating a level $k = 1$. Note that an equivalent interpretation is that we've found that the 2+1D bulk $\pi$-flux has a topological spin $e^{i\pi/4}$, corresponding to $\nu = 2$ in Kitaev's 16-fold way upon Higgs'ing from $U(1)$ to $\bZ_2$ \cite{KitaevAnyons}. The extension class $\nu_1(A)$ corresponds to the $\nu = 2$ mod 4 fusion rule for the vortices $a \otimes a = \epsilon$.

If we repeat the calculation for $U(1)_R$ we expect to find the inverse anomaly $k = -1$, since applying a parity-reversing transformation to the previous $U(1)_1$ coupled to $U(1)_L$ setup takes $U(1)_L$ to $U(1)_R$ and $U(1)_1$ to $U(1)_{-1}$. In the case of $U(1) = U(1)_R$ symmetry, we find that the charge is now
\begin{equation}Q = n/2 - m,\end{equation}
again with
\begin{equation}e^{2\pi i Q} = (-1)^s.\end{equation}
Now however
\begin{equation}Z_b(+/+,\theta/\theta) = \frac{1}{|\eta(\tau)|^2} \sum_{m,n \in \bZ} e^{ i \theta (n/2 + m - \theta/4\pi)} q^{\frac{1}{2} (n/2 + m + \theta/2\pi)^2} \bar q^{\frac{1}{2} (n/2 - m)^2}\end{equation}
so the $\theta$ flux has charge $-\theta/4\pi$, yielding the opposite anomaly $U(1)_{-1}.$ This corresponds to $\nu = -2$ in Kitaev's 16-fold way.

Note that we could've chosen another branching of the logarithm to define $Q$ in the twisted sectors, using
\begin{equation}-e^{i\theta r}\end{equation}
rather than $e^{i\theta r}$ (this amounts to a shift $\theta \mapsto \theta + 2\pi$). In terms of $Q$, this means
\begin{equation}Q = n/2 + m + n.\end{equation}
Because $s = n$, these different choices correspond to a symmetry of the anomaly theory:
\begin{equation}A \mapsto A + \pi B_1.\end{equation}
This does not change the Chern-Simons level, so the anomaly does not depend on our choices. This redefinition of the $U(1)$ symmetry in the bosonized theory maps to the same $U(1)$ symmetry in the fermionic theory because we will let $B_1$ be dynamical. This does not change the $\nu$'s, but later we will see there are automorphisms that do.

Finally, if we gauge $B_1$ without coupling to the spin factor $W$, we obtain a description of the Dirac fermion analogous to the usual bosonization of the Kitaev phases in Section \ref{s:kitaev}. In this case, by general reasoning having to do with the extension of $U(1)$ by $\bZ_2$ \cite{anomaliesvarious}, there is a mixed anomaly between $U(1)$ and the dual $\bZ_2$ symmetry, of the form
\begin{equation}\frac{1}{2} B_1' \frac{dA}{2\pi},\end{equation}
where $B_1'$ couples to the dual $\bZ_2$ symmetry, ie. the magnetic symmetry of $B_1$. The transmutation of an extension into a mixed anomaly is well-documented. See for instance \cite{anomaliesvarious,tachikawa2017gauging}.

We mention yet another perspective on the issue of bosonizing this chiral $U(1)$ symmetry also appeared in \cite{yuanyoshiki}.

\section{2nd bosonization and more general anomalies}

Now we wish to discuss the case that the fermionic SPT phase in the bulk is not of Gu-Wen-Freed type, instead described by $\nu_2,\nu_1,\nu_{0} \sim \nu_{-1}$. To do this, we will need to describe 2nd bosonization in the presence of a boundary. As before, the trick will be to express the configuration of Kitaev string and fermionic particle probes by Poincar\'e-Lefschetz duality. One finds that $\Sigma, \gamma$ are expressed by $C_1 \in C^{D-1}(X,\bZ_2), C_2 \in C^{D-2}(X,\bZ_2)$, $B_1 \in C^{D-2}(Y,\bZ_2), B_2 \in C^{D-3}(Y,\bZ_2)$ satisfying
\begin{equation}dC_2 = 0 \mod 2\end{equation}
\begin{equation}C_2|_Y = dB_2 \mod 2\end{equation}
\begin{equation}dC_1 = Sq^2 C_2 \mod 2\end{equation}
\begin{equation}C_1|_Y = \beta_1(0,B_2) + dB_1 \mod 2,\end{equation}
where $\beta_1(C_2,B_2)$ is a 1st descendant of the Postnikov class $Sq^2 C_2$ (expected to appear in Dirichlet boundary conditions because of its role in higher gauge transformations \cite{highersymm}), meaning
\begin{equation}d\beta_1(C_2,B_2) = Sq^2 (C_2 + dB_2) - Sq^2 C_2 \mod 2.\end{equation}
It's necessary to include it so that the last boundary condition is compatible with the Kitaev string Postnikov constraint \eqref{e:2ndbosconstraint}. Because the Steenrod squares are stable cohomology operations \cite{cohomologyoperations}, there is a choice of $\beta_1$ such that
\begin{equation}\beta_1(0,B_2) = Sq^2 B_2 \mod 2 \quad {\rm when} \quad dB_2 = 0 \mod 2.\end{equation}
This is very satisfying, since it shows that on the boundary in the absence of bulk probe fermions we get an $E_{2,D-1}$-gauge field $B = (B_1,B_2)$, as in the case of 1st bosonization. This pattern is expected to hold for all higher bosonizations because the groups of fermionic SRE phases are all abelian and form a spectrum whose Postnikov classes are all stable cohomology operations, which follows from \cite{KTTW, FreedHopkins}. See also \cite{xiong2018minimalist,gaiotto2017symmetry} for some general discussion on the spectrum structure of SRE phases, without assuming any connection to spin cobordism or TQFT.

To include a global $G$ symmetry, we make the replacements $C_2 \mapsto C_2 + \nu_2(A)$ and $C_1 \mapsto C_1 + C_2 \cup_{D-1} \nu_2(A) + \nu_1(A)$ as in Section \ref{s:morespts}, to find:
\begin{equation}\label{e:Gdirichlet}
d(C_2 + \nu_2(A)) = 0 \mod 2
\end{equation}
\begin{equation}(C_2 + \nu_2(A))|_Y = dB_2 \mod 2\end{equation}
\begin{equation}d(C_1 + C_2 \cup_{D-1} \nu_2(A) + \nu_1(A)) = Sq^2 (C_2 + \nu_2(A)) \mod 2\end{equation}
\begin{equation}(C_1 + C_2 \cup_{D-1} \nu_2(A) + \nu_1(A))|_Y = dB_1 + \beta_1(0,B_2) \mod 2.\end{equation}
When $C_2$ and $C_1$ are zero, meaning the probes lie only on the boundary, this becomes
\begin{equation}\label{e:2ndGextension}
d\nu_2(A) = 0 \mod 2
\end{equation}
\begin{equation}\nu_2(A)|_Y = dB_2 \mod 2\end{equation}
\begin{equation}d\nu_1(A) = Sq^2\nu_2(A) \mod 2\end{equation}
\begin{equation}\nu_1(A)|_Y = dB_1 + \beta_1(0,B_2) \mod 2.\end{equation}
Two of these are bulk constraints, and the other two describe a nontrivial extension of the boundary $G$ symmetry by the $(D-2)$-group $E_{2,D}$, as we saw in the Gu-Wen-Freed case. We expect this general structure holds for all higher bosonizations. Note that we have not included the twisting $\xi$, but once the twisted Postnikov constraints such as \eqref{e:2ndpostconstraintwisted} are fully understood, this will be a straightforward extension of our discussion.

\subsection{1+1D Chiral $\bZ_2$ Anomaly}\label{secmajanom}

\subsubsection{Overview}

Let us now consider the example of a massless free Majorana fermion in 1+1D. The fermion number of this theory is only conserved modulo 2, but like the Dirac fermion can be considered a sum of fermion parities from the left-moving and right-moving sectors. We study the anomalous $\bZ_2$ symmetries corresponding to the left-moving and right-moving chiral fermion parities, $(-1)^{F_L}$, $(-1)^{F_R}$. The anomaly theories of these symmetries are a bit harder to understand than the chiral $U(1)$ anomaly, because they are not Chern-Simons theories (or even spin-Dijkgraaf-Witten theories) and require 2nd bosonization to describe. Instead, it is known that the anomaly theory is nonabelian, related by bosonization to the Ising TQFT (odd $\nu$ in Kiteav's 16-fold way \cite{KitaevAnyons}). We will first verify this by studying the symmetry fluxes and then we will apply our bosonization map.

Let us denote the partition function of the left-movers on a torus with spin structure $\eta$ as $Z_L(\tau,\eta)$ and $Z_R(\bar \tau,\eta)$ for the right-movers. In terms of the $c = 1/2$ Virasoro characters $\chi_1, \chi_\epsilon,\chi_\sigma$ (with labels corresponding to the three chiral primaries of the Ising CFT), we have
\begin{equation}\label{e:spincharacters}
    Z_L(\tau,AP/AP) = \chi_1(\tau) + \chi_\epsilon(\tau)
\end{equation}
\begin{equation}Z_L(\tau,AP/P) = \chi_1(\tau) - \chi_\epsilon(\tau)\end{equation}
\begin{equation}Z_L(\tau,P/AP) = \sqrt{2}\chi_\sigma(\tau)\end{equation}
\begin{equation}Z_L(\tau,P/P) = 0,\end{equation}
and the right-mover partition functions are the complex conjugates. (The full partition functions are products of these.) These are spin characters in the sense that the $T$ transformation acts on them by permuting the spin structure as it does geometrically and then multiplying the characters by an overall phase\footnote{In fact, one can also check the $S$ matrix is also equivariantly diagonal on these partition functions. This is because the chiral Majorana spin-CFT is holomorphic, in that its conformal blocks are states in an \emph{invertible} 2+1D spin-TQFT. The same is true for the chiral Dirac spin-CFT.
}. Indeed:
\begin{equation}T:Z_L(\tau,AP/AP) \mapsto \chi_1(\tau) - \chi_\epsilon(\tau) = Z_L(\tau,AP/P)\end{equation}
\begin{equation}Z_L(\tau,AP/P) \mapsto \chi_1(\tau) + \chi_\epsilon(\tau) = Z_L(\tau,AP/AP)\end{equation}
\begin{equation}Z_L(\tau,P/AP) \mapsto e^{2\pi i/24}\sqrt{2}\chi_\sigma(\tau) = e^{2\pi i/24}Z_L(\tau,P/AP).\end{equation}

Denoting by $A_L$ a background gauge field for $(-1)^{F_L}$, and $A_R$ for $(-1)^{F_R}$ then the twisted partition function of the Majorana fermion in these backgrounds may be defined as
\begin{equation}Z_f(A_L,A_R,\eta) = Z_L(\tau,\eta + A_L)Z_R(\bar\tau,\eta+A_R).\end{equation}
In particular, on an AP/AP torus with $(-1)^{F_L}$ symmetry twist in the spatial cycle, we have
\begin{equation}\label{e:leftparityflux}
Z_f(-/+_L,AP/AP) = Z_L(\tau,P/AP)Z_R(\bar \tau,AP/AP) = \sqrt{2}\chi_\sigma(\tau)(\bar \chi_1(\tau) + \bar \chi_\epsilon(\tau)).
\end{equation}
Under a $T$ transformation $\tau \mapsto \tau+1$, this transforms to
\begin{equation}e^{2\pi i/16} \sqrt{2} \chi_\sigma(\tau)(\bar \chi_1(\tau) - \bar \chi_\epsilon(\tau)) = e^{2\pi i/16} Z_f(-/-_L,AP/P).\end{equation}
Note that the fact that the $T$ transformation does not fix the spin structure conspires with the sign change in $\chi_\epsilon$ from the fermi-odd states of $\bar \chi_\epsilon$ to make $Z_f$ modular-covariant up to the factor $e^{2\pi i/16}$. We ascribe this factor to the symmetry charge of the flux. If we place the theory at the boundary of the Ising TQFT with a $\sigma$ line in the flux sector, the topological spin of the $\sigma$ line precisely cancels this phase factor, yielding a modular covariant spin partition function. This topological spin corresponds to $\nu = 1$ in Kitaev's 16-fold way, hence a generator of $\Omega^3_{spin}(B\bZ_2) = \bZ_8$ by \cite{BGK}.

We also observe that had we used the right moving chiral fermion parity, we would have found
\begin{equation}\label{e:rightparityflux}
Z_f(-/+_R,AP/AP) = \sqrt{2}(\chi_1(\tau) + \chi_\epsilon(\tau))\bar \chi_\sigma(\tau)
\end{equation}
and hence derived the opposite phase factor, $e^{-2\pi i/16}$ from the $T$ transformation. Thus, the anomalies of the left and right chiral fermion parities are inverses, as expected. Indeed, if we have two massless Majorana fermions where the $\bZ_2$ symmetry acts on the left-movers of one species and the right-movers of the other species, then we can write a pairing between the species so that like charges share a mass term (ie. scattering interaction), symmetrically opening a trivial gap.

\subsubsection{0th, 1st, and 2nd Bosonization of the free Majorana}

Now let us consider bosonizing the massless Majorana fermion. It is well-known that the bosonized theory is the critical Ising model \cite{francesco2012conformal}. We can verify this easily using the 0th bosonization map for the torus and \eqref{e:spincharacters}:
\begin{equation}Z_b = \frac{1}{2}\bigg(Z_f(AP/AP) + Z_f(AP/P) + Z_f(P/AP) + Z_f(P/P)\bigg) = |\chi_1|^2 + |\chi_\epsilon|^2 + |\chi_\sigma|^2,\end{equation}
which we recognize as the torus partition function of the critical Ising model.

It remains to determine how $(B_1,B_2)$ and the chiral fermion parity act on this Ising CFT. First we will show that $B_1$ couples to the ordinary spin-flipping $\bZ_2$ symmetry of the Ising CFT. To do so, we use the 1st bosonization relation \eqref{e:1stGbos} to compute the torus partition function in the background with a $B_1$ twist around the temporal cycle:
\begin{equation}Z_b(+/-_{B_1}) = \frac{1}{2}\bigg( Z_f(AP/AP) + Z_f(P/AP) - Z_f(AP/P) - Z_f(P/P) \bigg)\end{equation}\begin{equation} = |\chi_1|^2+|\chi_\epsilon|^2-|\chi_\sigma|^2.\end{equation}
We see the $\sigma$ sector, corresponding to the Ising spin operator, has odd charge while the $1,\epsilon$ sectors have even charge. Thus, $B_1$ couples to the spin-flip symmetry.
% Equivalently, the $B_1$ domain wall is the $\epsilon$-defect in the Ising CFT \cite{FUCHS,BGK}.

Next, to see how $B_2$ couples, we use the 2nd bosonization relation \eqref{e:2ndbos}. Recall we have $B_2 \in Z^0(X,\bZ_2)$, so the $B_2$ background is just a constant 0 or 1. If $B_2 = 0$, the partition functions are unmodified. On the other hand if $B_2 = 1$, then we obtain from the spin factor
\begin{equation}W_2(\eta,0,1) = {\rm Arf}(\eta),\end{equation}
which is $+1$ for all torus spin structures except for $P/P$, for which it is $-1$ (cf. \eqref{eqn2dspinfactor}). However, because $Z_f(P/P) = 0$, this factor does not affect the torus partition functions so it looks like a trivial modification of the theory to include this factor.

Actually, $B_2 = 1$ represents a transformation to a dual set of variables in this theory. If we then turn on a mass term for the Majorana fermion, the presence of $W_2(\eta,0,1)$ in the path integral swaps the gapped phases at $m>0$ and $m<0$. It is known that in bosonization the mass deformation of the Majorana fermion corresponds to tuning the relative strength of the two competing terms in the Ising Hamiltonian, so that the neighboring gapped phases bosonize to the ferromagnetic and paramagnetic gapped phases of the Ising chain.

Let us show this from our perspective, by forming the mass deformation to the trivial phase of the Majorana fermion, for which the partition function may be normalized so $Z_f(\eta) = 1$. Writing the analog of the bosonization relations \eqref{e:2ndbos} for the partition functions, suppressing normalization, we have
\begin{equation}Z_b(B_1,B_2) = \sum_\eta Z_f(\eta) W_2(\eta,B_1,B_2)= \sum_\eta W_2(\eta,B_1,B_2).\end{equation}
There are two cases:
\begin{equation}Z_b(B_1,B_2 = 1) = \sum_\eta W_2(\eta,B_1,B_2 = 1) = 1,\end{equation}
\begin{equation}Z_b(B_1,B_2 = 0) = \sum_\eta W_2(\eta,B_1,B_2 = 0) = \delta(B_1).\end{equation}
The first corresponds to the paramagnetic phase (the domain wall is massless and $B_1$ is deconfined) while the second corresponds to the ferromagnetic phase (the domain wall is massive and $B_1$ is confined). Thus we see that for a particular choice of relevant deformation of the 2nd bosonization, changing the value of $B_2$ changes which of the two neighboring gapped phases this deformation flows to. In this way, taking $B_2 = 1$ amounts to choosing a different ``duality frame" in the bosonized theory. We discuss this in more generality in Section \ref{s:groupoids}.

% we can use the known transformation rules of $c = 1/2$ conformal characters to show that this transforms to
% \begin{equation}Z_f(-/+_L,AP/AP) \mapsto e^{2\pi i/24} Z_L(\tau,P/AP) e^{2\pi i/48} Z_R(\bar \tau,AP/P) = e^{2\pi i/16} Z_f(-/-_L,AP/P),\end{equation}
% yielding the topological spin of the flux, which we interpreted in the previous section as its symmetry charge. We see that this corresponds to Kitaev's $\nu = 1$ in the 16-fold way \cite{KitaevAnyons}, hence a generator of $\Omega^3_{spin}(B\bZ_2) = \bZ_2$ by \cite{BGK}. On the other hand, for $(-1)^{F_R}$, we have
% \begin{equation}Z_f(-/+_R,AP/AP) = Z_L(\tau,AP/AP)Z_R(\bar \tau,P/AP),\end{equation}
% which is the complex conjugate of $Z_f(+/-_L,AP/AP)$, yielding an opposite topological spin for the flux, hence $\nu = -1$ and the inverse anomaly theory.

\subsubsection{Bosonized Chiral Anomaly}\label{secchiralbos}

Now we consider the anomaly in 2nd bosonization. We can present the SPT phases associated with the left and right chiral symmetries as
\begin{equation}\label{e:majbosdata}
    \nu_2 = A \quad \nu_1 = 0 \quad \nu_0 = 0,
\end{equation}
\begin{equation}\nu_2 = A \quad \nu_1 = A^2 \quad \nu_0 = - \frac{1}{4} A dA,\end{equation}
respectively. See \cite{BGK} for a discussion of the group law this data satisfies, which also shows that they are inverses. Let us check that these values correctly predict the modified symmetry algebra.

First, we will show
\begin{equation}\nu_2 = A\end{equation}
for either symmetry. This is done by computing
\begin{equation}Z_b(+/-_{L,R}) = |\chi_1|^2 - |\chi_\epsilon|^2.\end{equation}
We see that in bosonization, either chiral fermion parity takes $\epsilon \to -\epsilon$, $\sigma \to 0$. This is characteristic of Kramers-Wannier duality, since $\epsilon$ perturbs the critical Ising chain to the symmetry breaking or trivial phase, depending on its sign. There is no such symmetry of the critical Ising chain which realizes this duality, since it maps the local operator $\sigma$ to the Kadanoff disorder operator, which occurs at the end of a string operator. Thus, to couple the critical Ising chain to the $A$ background, we need to apply $B_2$ such that
\begin{equation}\label{e:majoranaconstraint}
    dB_2 = A \mod 2
\end{equation}
From this by comparing with \eqref{e:2ndGextension}, we find $\nu_2 = A$ for either symmetry.

It remains to compute $\nu_1(A)$. To do this, we have to study the chiral fermion parity domain wall. It is well-known \cite{CALLAN1985427} that this domain wall supports a Majorana zero mode. When two domain walls fuse together, the two zero modes are collected into a single Dirac Fock space which has two basis states with opposite fermion parity, the occupied and un-occupied states. Fusion vertices for symmetry defects need to be non-degenerate, so the symmetry chooses one of the states in this Fock space. Furthermore, they must fuse to a state of definite fermion parity, since fusion is local. Thus, there are two fusion channels: bosonic and fermionic.

When the fusion results in the bosonic state, we have $\nu_1(A) = 0$, and when it results in the fermionic state, we have $\nu_1(A) = A^2$. Indeed, in the latter case, after we sum over spin structures the $B_1$ domain wall is bound to this fermionic particle, so we obtain a domain wall junction $A \otimes A = B_1$, indicating a nontrivial extension of $G$ by $\bZ_2$.

An important caveat, however is that whether $\nu_1(A_L) = 0$, $\nu_1(A_R) = A_R^2$ or whether it is the opposite is ultimately a matter of convention in how one formulates the SPT phase. Indeed, the AHSS has a $\bZ_4$ automorphism in 2+1D:
\begin{equation}\nu_2 \mapsto \nu_2\end{equation}
\begin{equation}\nu_1 \mapsto \nu_1 + Sq^1 \nu_2,\end{equation}
\begin{equation}\nu_0 \mapsto \nu_0 - \frac{1}{4} \nu_2 d\nu_2,\end{equation}
which amounts to a field redefinition
\begin{equation}C_1 \mapsto C_1 + Sq^1 C_2\end{equation}
and exchanges \eqref{e:majbosdata}. We are allowed to do this because in describing the fermionic theory $C_1$ and $C_2$ are dynamical, and this is a redefinition of the sum. One can think of this $\bZ_4$ as the Galois action on the 8 root phases of Kitaev's 16-fold way. This contrasts with the $U(1)$ case where there was no such automorphism. We note that this automorphism does not persist to higher dimensions, as $Sq^1 Sq^2 Sq^1 C_2 = Sq^3 Sq^1 C_2$ is no longer zero when the degree of $C_2$ is more than 1. It would be interesting to understand in general what this ``Galois group" is.

However, we can at least show that $\nu_1$ have to be opposites between the two chiral fermion parities. For instance, suppose two $(-1)^{F_L}$ domain walls with Majorana zero modes $c_1,c_2$, labelled from left to right, come into contact, pairing by a Hamiltonian term
\begin{equation}i c_1 c_2.\end{equation}
Then, applying reflection symmetry, these become $(-1)^{F_R}$ domain walls now pairing by the opposite term
\begin{equation}- i c_1 c_2,\end{equation}
since $c_1$ and $c_2$ anti-commute. Thus, the ground states of these pairing terms have opposite fermion parity.

Let us give another argument. Write $P_{L,R}$ as the chiral fermion parity operators. Suppose $P_L^2 = 1$. We have $P_R = (-1)^F P_L$ and we're interested in
\begin{equation}P_R^2 = (-1)^F P_L (-1)^F P_L.\end{equation}
We will argue that $(-1)^F$ and $P_L$ anti-commute on the spatial circle with periodic spin structure. This can be seen by studying the mass deformations of the Majorana fermion, which are distinguished by the fermion parity of their ground states on the periodic circle. Actually since their $P/P$ torus partition functions differ by an over-all sign for fixed absolute value of the mass, all of their corresponding states have opposite fermion parity. Since $P_L$ switches the sign of the mass, it therefore anti-commutes with $(-1)^F$ in these sectors (the same argument shows $P_R$ has the same anti-commutation relation with $(-1)^F$, which is required for consistency). Thus we have
\begin{equation}P_R^2 = -1\end{equation}
in the periodic sector. Recalling the definition of $W(\eta,B_1)$, this is precisely the sector of $B_1$-odd states. Thus if $P_L^2 = 1$, then $P_R^2$ is the spin flip symmetry.

\section{Time Reversal Anomalies in 2+1D}

We will comment on our expectations for time reversal $T^2 = (-1)^F$ anomalies of 2+1D fermionic systems in bosonization. This is done by analogy with the above discussions, although plenty of detail is missing. To complete the story will require a deeper understanding of bosonization in 3+1D than has been so far achieved, mostly because of difficulties in understanding the generalized Gu-Wen-Freed equation. This leaves $\nu_0$ a mystery. However, we can describe $\nu_3,\nu_2,\nu_1$ and how they modify the time reversal symmetry algebra of the anomalous theory upon bosonization. This is enough to determine such anomalies up to a bosonic anomaly (namely $w_1(TX)^4 \in \Omega^4_O$).

Recall that such anomalies (equivalently 3+1D topological superconductors) are classified by
\begin{equation}\Omega^4_{spin}(B\bZ_2, \xi) = \bZ_{16},\end{equation}
where $\xi$ is a sum of three copies of the sign representation $\sigma$ \cite{KTTW}. This group is computed by the AHSS as an iterated extension \cite{kitaevIpam} of
\begin{equation}H^1(B\bZ_2,\bZ^{\xi}) = \bZ_2\end{equation}
\begin{equation}{\rm by\ } H^2(B\bZ_2, \bZ_2) = \bZ_2\end{equation}
\begin{equation}{\rm by\ } H^3(B\bZ_2, \bZ_2) = \bZ_2\end{equation}
\begin{equation}{\rm by\ } H^5(B\bZ_2, \bZ^{\xi}) = \bZ_2.\end{equation}
Thus we expect the root phases ($\nu$ odd) to have
\begin{equation}\nu_3(A) = A \in H^1(B\bZ_2,\bZ^{\xi}).\end{equation}
This means that the time reversal ``domain wall" carries an odd number of $p+ip$ superconductors. This is what we know from the boundary of the odd $\nu$ topological superconductors, that breaking $T$ in two different ways on the boundary produces a domain wall carrying a $c = 1/2 \mod 1$ chiral mode, which can be thought of as the boundary of a $p+ip$ superconductor which forms the bulk domain wall \cite{FCV}.

Analogous to the situation for the Majorana fermion in 1+1D, we see that in bosonization, we'll have
\begin{equation}d_G B_3 = dB_3 - 2 A \cup B_3 = \nu_3(A) = A,\end{equation}
which implies $A = dB_3$ mod 2, so we cannot place the bosonized theory on an unorientable manifold. We expect that this means that time reversal symmetry is realized as an anti-unitary duality of the bosonized theory. This is a concrete prediction, since the boundary of the $\nu = 1$ topological superconductor may be realized by a massless free Majorana fermion in 2+1D. We expect that the bosonization has an anti-unitary ``self-duality" inherited from time reversal of the Majorana fermion, which becomes a symmetry once we gauge the $E_{2,3}$ 2-group symmetry.

One such bosonization of the 2+1D massless free Majorana (which is expected to at least hold in the IR) is $SU(2)_2$ Chern-Simons theory coupled to a massless adjoint scalar $\phi$ \cite{majoranaduality}\footnote{In the reference, they studied $SO(3)_1$, which is obtained from $SU(2)_2$ by gauging the center symmetry.}. We expect the $B\bZ_2$ symmetry coupling to $B_1$ is the center symmetry of the gauge field and $B_2$ couples to the $\bZ_2$ symmetry $\phi \mapsto -\phi$. This bosonization realizes the Majorana as the critical point between the $\nu = 3$ ($SU(2)_2$) and $\nu = 4$ (double semion) topological theories in Kitaev's 16-fold way \cite{KitaevAnyons}. When we take $B_3 \in Z^0(X,\bZ)$ to be a constant $m \in \bZ$, it means we apply $m$ $p+ip$ superconductors and then bosonize (see the discussion around $B_2$ for the 1+1D Majorana). This changes the bosonization to one realized as a critical point between phases $3 + m$ and $4 + m$ in the 16-fold way.

We expect all of these theories are dual by various combinations of particle-vortex/level-rank dualities \cite{HsinSeiberg}. $\nu_3(A) = A$ indicates that time reversal must be composed with such a duality in bosonization. Indeed, it takes $SU(2)_2 + \phi$ to $SU(2)_{-2} + \phi$, which realizes a critical point between $\nu = -4$ and $\nu = -3$ and we must compose with a certain duality to return to a transition from $\nu = 3$ to $\nu = 4$. When we gauge the $E_{2,3}$ symmetry of the bosonization, this duality becomes a symmetry. It would be very interesting to understand this in more detail.

For anomalies with even $\nu$ mod 16, we will have $\nu_3 = 0$. It will be possible to place the bosonizations of these theories on unorientable spacetimes, ie. we expect the bosonized time reversal to act as a symmetry, not a duality. However, $(\nu_1,\nu_2)$ shall describe how time reversal is nontrivially extended by $E_{2,3}$. In particular, $\nu_2 \in H^2(B\bZ_2,\bZ_2)$ denotes a group extension of the ordinary sort, while $\nu_1$ denotes an extension of this extended group by the 1-form symmetry 2-group $B\bZ_2$ (a Postnikov class for time reversal). Both phenomena have already been encountered in studying time reversal symmetry of gapped phases in 2+1D \cite{maissam}.

See also \cite{kobayashi2019pin} for another recent approach to anomalies in this symmetry class and dimension, similar to what we outline here, as well as \cite{kobayashi2019gapped}. For field theory approaches, see \cite{witten2016parity,cordova2018time}.

\section{Some Comments on Symmetry Groupoids}\label{s:groupoids}

We have studied how the 1+1D free Majorana couples to a $E_{2,2}$ background
\begin{equation}B_1 \in Z^1(X,\bZ_2)\end{equation}
\begin{equation}B_2 \in Z^0(X,\bZ_2).\end{equation}
The first piece, $B_1$ is easily understood as a gauge field which couples to a $\bZ_2$ symmetry, but how can we understand $B_2$? Likewise in 2+1D, $B_3 \in Z^0(X,\bZ)$. What are the meaning of these 0-cocycles as ``0-form gauge fields", which on connected spacetimes are just constant functions?

\begin{figure}
    \centering
    \includegraphics{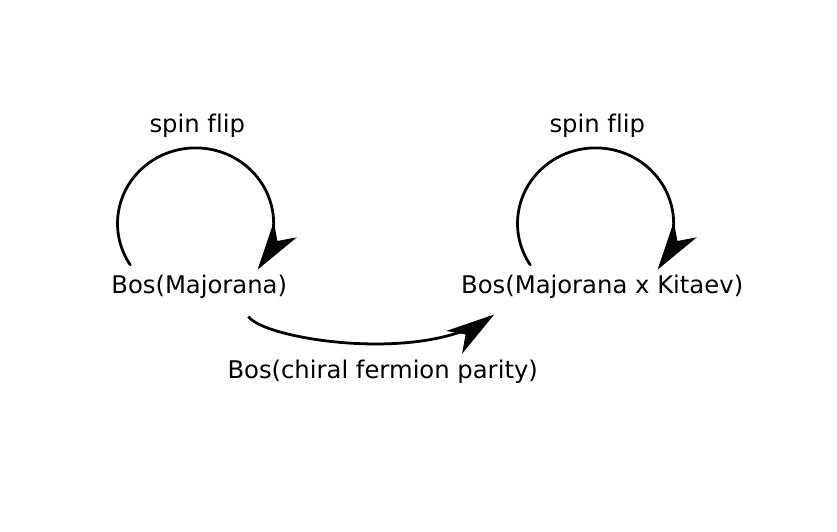}
    \caption{The $G = \bZ_2$ symmetry action of the chiral fermion parity of the free massless Majorana fermion in 1+1D when bosonized is extended nontrivially by the canonical $E_{2,2}$ groupoid symmetry of 1+1D bosonizations, such that the generator of $G$ acts as a duality transformation between two a priori inequivalent bosonizations.}
    \label{fig:my_label}
\end{figure}

We propose that the proper way to think about $E_{2,2}$ and $E_{3,3}$ are as groupoids, which is like a group with multiple base points \cite{groupoids}. Precisely it is a category whose morphisms are all invertible. Simple groupoids like $E_{2,2}$ can be visualized graphically, as in Fig. \ref{fig:my_label}.

Groupoids typically do not act\footnote{Concisely, a groupoid action on a category is a functor from the groupoid in question to that category. For a description of the category of QFTs see \cite{kapustinicm}.} faithfully on individual quantum field theories. Instead, a groupoid acts on a collection of QFTs $T(x)$, indexed by the objects $x$ of the groupoid. For each morphism $f:x \to y$ in the groupoid we have a Hilbert space transformation (either unitary or anti-unitary)
\begin{equation}U(f): \cH_{T(x)} \to \cH_{T(y)}\end{equation}
such that if $H(x), H(y)$ are the Hamiltonians of $T(x), T(y)$, then
\begin{equation}U(f) H(x) = H(y) U(f).\end{equation}
When $x = y$, this means that the automorphisms of $x$ act as ordinary symmetries of $T(x)$. Indeed, when the groupoid has a single object, it is equivalently just a group and this notion of groupoid symmetry specializes to the usual notion of group symmetry. However, when $x \neq y$, $U(f)$ describes a \emph{duality} between $T(x)$ and $T(y)$. In this way, groupoid symmetry generalizes both ordinary group symmetry and duality. Note that $E_{2,2}$ doesn't contain any dualities but also isn't a group.

The upshot of this is that we bosonize a 1+1D fermionic QFT, we obtain not one bosonic QFT, but \emph{two} bosonic QFTs, depending on whether $B_2 = 0$ or $1$ (we assume $X$ is connected), each with an action of $\bZ_2$. The theory obtained from $B_2 = 0$ is the ordinary 1st bosonization \eqref{e:1stbos}, while the theory obtained from $B_2 = 1$ is the 1st bosonization after tensoring with the unique nontrivial 1+1D FSRE phase, the Kitaev wire. One way to summarize this is to say that $E_{2,2}$ acts on the bosonizations of any 1+1D fermionic QFT.

In general dimensions we will get several bosonizations the same way by tensoring with the different FSRE phases in that dimension. For instance, in 2+1D we can define infinitely many bosonizations by tensoring with different numbers of $p+ip$ superconductors, which constitute the group of 2+1D FSRE phases \cite{KTTW}. In general, $E_{D,D}$ will act naturally on these bosonizations, reflecting the fact that $E_{D,D}$ is a (higher) groupoid whose components are labelled by the different $D$-dimensional FSRE phases.

Recall when we studied the chiral $\bZ_2$ symmetry of the 1+1D massless Majorana fermion we found the constraint \eqref{e:majoranaconstraint}:
\begin{equation}dB_1 = \nu_1(A) = 0\end{equation}
\begin{equation}dB_2 = \nu_2(A) = A.\end{equation}
We can interpret this as describing an extension of $G = \bZ_2$ by $E_{2,2}$ as groupoids \cite{groupoids}. The extension $\hat G$ is the groupoid shown in Fig. \ref{fig:my_label}. When we look at the action of $\hat G$ on the bosonizations we see that the generator of $G$ lifts to a Kramers-Wannier duality between the two Ising chains. The fact that $\nu_1 = 0$ implies that when we perform this operation twice (dualizing $A \to B \to A$), we obtain the identity transformation. If $\nu_1(A)$ was nontrivial instead, doing this twice would instead yield the $\bZ_2$ symmetry operator.

We can put it another way in $1+1D$. It is known in conformal field theory that the conformal defects also can capture dualities, such as in \cite{FUCHS}. These defects $X$ can be grouplike, meaning $X \otimes X^\vee$ is the unit defect, corresponding to the domain walls of ordinary global symmetries, where $X^\vee$ is the dual defect to $X$. They can also be duality defects, meaning that $X \otimes X^\vee$ is a sum of grouplike defects. The sort of groupoid-extended $G$ action we're talking about assigns either a grouplike or duality defect $X(g)$ to every element $g \in G$, which respects the fusion rules, meaning $X(gh) = X(g) \otimes X(h)$. Further, it includes a choice of grouplike defect in $X(g) \otimes X(h) \otimes X(gh)^\vee = X(gh) \otimes X(gh)^\vee$ (this is $\nu_1$). This defines a group extension of $G$ by the grouplike defects (this is $\nu_2$). In higher dimensions we expect a similar structure to be definable.

\appendix
\addcontentsline{toc}{section}{Appendices}
\section*{Appendices}
% \section{Lattice Bosonization with Boundary}

% In this appendix, we describe an extension of the generalized Jordan-Wigner mappings of \cite{CKR,CK} to spaces with boundary. For simplicity, we describe this only for triangulated spaces.

% Let $X$ be a triangulated $n$-manifold with boundary $\partial X = Y$. We define the algebra of fermionic operators associated to $X$ to be the Clifford algebra generated by two species of Majorana operators $\gamma_\sigma, \gamma_\sigma'$ per $n$-simplex $\sigma$, each Hermitian and satisfying $\gamma^2 = 1$ and anticommuting with the other generators.

% Let us choose a branching structure on $X$ such that all the $n$-simplices of $X$ are ordered and all edges in the interior of $X$ ending at vertices of $Y$ point into $X$.

% A particularly simple branching structure exists on a barycentric subdivision of $X$... USE DISCRETE MORSE THEORY

% The key geometrical observation is that a branching structure on a triangulated space $X$ restricts to a branching structure on its boundary $Y:= \partial X$. If $X$ is an $n$-manifold, it can be shown \cite{thesis} that the branching structure defines an $n-2$-cycle $W_2(X) \in Z_{n-2}(X,\bZ_2)$, and an $n-3$-chain $W_2(TY)$

\section{Derivation of the Twisted Gu-Wen-Freed Equation}\label{s:gu-wen-freed}

In this section, we derive the twisted Gu-Wen-Freed equation \eqref{e:gu-wen-freed-twist}. We find the same constraint that was recently derived in \cite{LZW}. A version of the constraint also appeared in the 2+1D Pin$^+$ case in \cite{bhardwaj}. All of the results are so far consistent. This derivation is equivalent to deriving the differential
\begin{equation}d_2:H^{D-1}(X,\Omega^1_{spin}) \to H^{D+1}(X,U(1)^\xi)\end{equation}
in the Atiyah-Hirzebruch spectral sequence.

First of all, by naturality of the spectral sequence, and because it is the first (possibly) nonvanishing differential, this has to be a stable cohomology operation, meaning $d_2 \nu_1$ is linear in $\nu_1 \in H^{D-1}(X,\Omega^1_{spin})$. We already know in the untwisted case
\begin{equation}d_2 \nu_1 = \frac{1}{2} Sq^2 \nu_1 \qquad \xi = 0,\end{equation}
as has been derived in \cite{GuWen,freed2008}. In the general twisted case,
\begin{equation}d_2 \nu_1 = \frac{1}{2} Sq^2 \nu_1 + \frac{k_1}{2} w_2 \nu_1 + \frac{k_2}{2} w_1^2 \nu_1,\end{equation}
for some universal $k_1, k_2 \in \bZ_2$ we need to determine. We can do this by a couple of well-chosen examples. Note that a third possible term $\frac{k_3}{2} w_1 Sq^1 \nu_1$ is actually exact.

\subsection{$\Omega^2_{pin-} = \bZ_8$, $k_2 = 0$}

The first symmetry class we study is time reversal $T^2 = 1$ in 1+1D. This corresponds to $\Omega^2_{pin-} = \Omega^2_{spin}(B\bZ_2,\xi)$, where $w_1(\xi) = A$, $w_2(\xi) = 0$, and $A$ is the $\bZ_2$ gauge field. There is a $\bZ_8$ of such phases, generated by the Kitaev wire \cite{FidkowskiKitaev}.

The $E_2$ page of the AHSS has
\begin{equation}H^2(B\bZ_2,U(1)^\xi) = \bZ_2\end{equation}
\begin{equation}H^1(B\bZ_2,\Omega^1_{spin}) = \bZ_2\end{equation}
\begin{equation}H^0(B\bZ_2,\Omega^2_{spin}) = \bZ_2,\end{equation}
so all differentials have to vanish. Meanwhile,
\begin{equation}d_2 \nu_1 = \frac{k_2}{2} A^2 \nu_1.\end{equation}
Thus, we find $k_2 = 0$.

\subsection{$\Omega^1_{pin+} = 0$, $k_1 = 1$}\label{s:1pin+}

Now we study 0+1D systems with $T^2 = (-1)^F$. This corresponds to $\Omega^1_{pin+} = \Omega^1_{spin}(B\bZ_2,\xi)$ with $w_1(\xi) = A, w_2(\xi) = A^2$, and $A$ is the $\bZ_2$ gauge field. There are no such phases \cite{KTTW}.

The $E_2$ page of the AHSS has
\begin{equation}H^1(B\bZ_2,U(1)^\xi) = 0\end{equation}
\begin{equation}H^0(B\bZ_2,\Omega^1_{spin}) = \bZ_2.\end{equation}
Thus, the $d_2$ differential has to eat the possible Gu-Wen-Freed phase. We have
\begin{equation}d_2 \nu_1 = \frac{k_1}{2} A^2 \nu_1 \neq 0.\end{equation}
It follows $k_1 = 1$. This concludes our derivation of the twisted Gu-Wen-Freed equation.

\section{Derivation of the 2nd Bosonization Postnikov Constraint}\label{s:twisted-postnikov}

In this section we study the first differential in the 2+1D AHSS from the Kitaev string part:
\begin{equation}d_2:H^{D-2}(X,\Omega^2_{spin}) \to H^{D}(X,\Omega^1_{spin})\end{equation}
which controls the Postnikov constraint \eqref{e:2ndpostconstraintwisted}. By naturality of the spectral sequence, and because it is the first (possibly) nonvanishing differential, this has to be a stable cohomology operation, meaning $d_2 \nu_2$ is linear in $\nu_2 \in H^{D-2}(X,\Omega^2_{spin})$.

We know that in the untwisted ($\xi = 0$) case \cite{higherbos} that
\begin{equation}d_2 \nu_2 = Sq^2 \nu_2 \qquad \xi = 0.\end{equation}
With a general twist, the most general stable cohomology operation we can write down is
\begin{equation}d_2 \nu_2 = Sq^2 \nu_2 + k_1 w_2(\xi) \nu_2 +  k_2 w_1(\xi)^2 \nu_2 + k_3 w_1(\xi) Sq^1 \nu_2,\end{equation}
where $k_1, k_2, k_3 \in \bZ_2$ are universal constants (independent of $D$) to be determined. Because of the universality of this expression, we need only consider a few examples.

\subsection{$\Omega^3_{spin}(B\bZ_2,2\sigma) = 2\bZ$, $k_1 = 1$}

First we consider the symmetry class of a unitary, orientation preserving symmetry $C$ with $C^2 = (-1)^F$ in 2+1D. It is known that the only phase in this symmetry class is a stack of two $p+ip$ superconductors\footnote{A single $p+ip$ has an anomaly. Its action is defined using a 4-manifold filling using the signature form divided by 16 \cite{KTTW}. With a $\bZ_2$ symmetry background with $C^2 = (-1)^F$, called a spin-$\bZ_2$ structure, the 4-manifold is taken to have spin-$\bZ_2$ structure extending that on the 3-manifold. Closed 4-manifolds with this structure measure possible ambiguity, and since the Enriques complex surface is a 4-manifold with signature 8 and spin-$\bZ_2$ structure, we can only consistently use the signature form divided by 8, which is the partition function of two $p+ip$ superconductors.}. This symmetry class has $w_1(\xi) = 0, w_2(\xi) = A^2$, where $A$ is the $\bZ_2$ gauge field.

In particular, this means that the possible Gu-Wen-Freed phase in this dimension, coming from
\begin{equation}\nu_1 = A^2 \in H^2(B\bZ_2,\Omega^1_{spin})\end{equation}
has to be eaten by a differential. There are two possibilities: either $d_2 \nu_1 \neq 0$ (meaning $\nu_1$ fails the twisted Gu-Wen-Freed equation) or $\nu_2 = d_2 \alpha$ for some $\alpha \in H^0(B\bZ_2,\Omega^2_{spin})$. We have
\begin{equation}d_2 \nu_1 \in H^4(B\bZ_2,U(1)) = 0\end{equation}
regardless of $k_2$, so it must be the latter. $H^0(B\bZ_2,\Omega^2_{spin}) = \bZ_2$ generated by a constant 0-chain $\mathbbm{1}$, and we have
\begin{equation}d_2 \mathbbm{1} = k_1 A^2.\end{equation}
Therefore, for $d_2 \alpha = \nu_2$, we must have $k_1 = 1$.

A physical interpretation of this result is that the boundary of the Gu-Wen-Freed phase has an instanton $\nu_1 = A^2$ where fermion parity is not conserved. However, if we stack a Kitaev string on the boundary, then because of the twist, this instanton acts as a spin structure defect, and the ground state of the Kitaev string flips fermion parity as it passes this defect, fixing the fermion parity conservation.

We note that this differential also eats the possible Kitaev string phase in this symmetry class and dimension, which would have $\nu_2 = A \in H^1(B\bZ_2,\Omega^2_{spin})$. Indeed, since $k_2 = 1,$
\begin{equation}d_2 \nu_2 = A^3 \neq 0 \in H^3(B\bZ_2,\Omega^1_{spin}).\end{equation}
This is an equally good way to derive $k_1 = 1$ and it makes for a nice sanity check.

\subsection{$\Omega^2_{pin+} = \bZ_2$, $k_2 = 0$}

Next we consider the symmetry class $T^2 = (-1)^F$ in 1+1D. This corresponds to $\Omega^2_{pin+} = \Omega^2_{spin}(B\bZ_2,\xi)$, where $w_1(\xi) = A$, $w_2(\xi) = A^2$, and $A$ is the $\bZ_2$ gauge field. It is known that there is one such phase, and it is a Gu-Wen-Freed phase with $\nu_1 = A$ \cite{FidkowskiKitaev,KTTW}.

The $E_2$ page of the AHSS has
\begin{equation}H^2(B\bZ_2,U(1)^\xi) = \bZ_2\end{equation}
\begin{equation}H^1(B\bZ_2,\Omega^1_{spin}) = \bZ_2\end{equation}
\begin{equation}H^0(B\bZ_2,\Omega^2_{spin}) = \bZ_2.\end{equation}
The bosonic symmetry class is eaten by the $d_2$ differential from the anomalous Gu-Wen-Freed phase in 1+1D we discussed in \ref{s:1pin+}. The Gu-Wen-Freed phase with $\nu_1 = A$ lives because
\begin{equation}d_2 \nu_1 \in H^3(B\bZ_2,U(1)^\xi) = 0.\end{equation}
This phase accounts for the entire group of SPT phases in this symmetry class and dimension.

Therefore, the possible Kitaev string phase with $\nu_2 = \mathbbm{1}$ must not contribute. It cannot be the target of a differential so it has to be eaten by either $d_2$ or $d_3$. However, $d_3$ lands in an empty cohomology group, so we must have
\begin{equation}d_2 \mathbbm{1} = (1 + k_2) A^2 \neq 0,\end{equation}
which implies $k_2 = 0$.

\subsection{$\Omega^3_{pin-} = 0$, $k_3 = 1$}

Next we consider time reversal symmetry $T$ in 2+1D with $T^2 = 1$. It is known that there are no phases in this symmetry class \cite{KirbyTaylor}. The relevant cobordism group is $\Omega^3_{pin-} = \Omega^3_{spin}(B\bZ_2,\xi)$ where $w_1(\xi) = A$, $w_2(\xi) = 0$, and $A$ is the $\bZ_2$ gauge field.

First of all, there are no bosonic phases, since $H^3(B\bZ_2,U(1)^\xi) = 0$. Indeed, the generator without the twist, $\frac{1}{2} A^3$ is actually eaten by the twisted differential (cf. Eq \eqref{e:twisteddiff}):
\begin{equation}d_A \frac{1}{4} A^2 = \frac{1}{4}(dA^2 - 2 A^3) = \frac{1}{2}A^3.\end{equation}

Next, the Gu-Wen-Freed phases in $H^2(B\bZ_2,\Omega^1_{spin}) = \bZ_2$ have a possible generator $\nu_1 = A^2$. However, this generator fails the Gu-Wen-Freed equation \eqref{e:gu-wen-freed-twist}, serving as a boundary counterterm to the bosonic 3+1D TRS phase $w_1^4$.

Finally, there is a possible Kitaev string phase with $\nu_2 = A$. This cannot be the target of a differential, and must instead be eaten either by $d_2$ or $d_3$. However, the only class $d_3$ can land on was already eaten by $d_2$ of the possible Gu-Wen-Freed phase above. Thus, it must be eaten by $d_2$. We have
\begin{equation}d_2 \nu_2 = k_3 A^3,\end{equation}
from which we conclude $k_3 = 1$. This concludes the derivation of the 2nd bosonization Postnikov constraint.

\section{A Derivation of the 3rd Bosonization Postnikov Constraint}\label{s:3rdpostder}

In this section we derive \eqref{e:3rdpostconstraint} by studying the first differential in the 2+1D AHSS from the $p+ip$ membrane part:
\begin{equation}d_2:H^{D-3}(X,\Omega^3_{spin}) \to H^{D-1}(X,\Omega^2_{spin}).\end{equation}
Note that our spin cobordism is the Anderson dual of the usual spin bordism, so that $\Omega^3_{spin} = \bZ^\xi$. Since $d_2$ is the first non-vanishing differential, it is linear in $\nu_3$, so it takes the form
\begin{equation}d_2 \nu_3 = k_1 Sq^2 \nu_3 + k_2 w_2(\xi) \nu_3 + k_3 w_1(\xi)^2 \nu_3,\end{equation}
where $k_1, k_2, k_3 \in \bZ_2$ are universal constants to be determined. We note that a forth possible term $w_1(\xi) Sq^1 \nu_3$ is actually equivalent to $w_1(\xi)^2 \nu_3$ because $\nu_3$ is a $\bZ^\xi$-cocycle. As in the previous appendices, we determine $k_1, k_2, k_3$ by some well-chosen examples.

\subsection{$\Omega^4_{pin+} = \bZ_{16}, k_2 = k_3$}

First we consider the symmetry class $T^2 = (-1)^F$ in 3+1D, corresponding to $\Omega^4_{pin+} = \Omega^4_{spin}(B\bZ_2,\xi)$ with $w_1(\xi) = A$, $w_2(\xi) = A^2$, where $A$ is the $\bZ_2$ gauge field. These are the usual 3+1D topological superconductors, and there is a $\bZ_{16}$ classification of them \cite{KTTW}.

The $E_2$ page of the AHSS has
\begin{equation}H^4(B\bZ_2,U(1)^\xi) = \bZ_2\end{equation}
\begin{equation}H^3(B\bZ_2,\Omega^1_{spin}) = \bZ_2\end{equation}
\begin{equation}H^2(B\bZ_2,\Omega^2_{spin}) = \bZ_2\end{equation}
\begin{equation}H^1(B\bZ_2,\Omega^3_{spin}) = \bZ_2,\end{equation}
so to obtain 16 phases, all differentials must vanish. In particular, this means that the root phases with $\nu_3 = A$ we must have
\begin{equation}d_2 A = k_2 A^3 + k_3 A^3 = 0,\end{equation}
from which we find $k_2 + k_3 = 0$.

\subsection{$\Omega^4_{pin-} = 0, k_3 = 1$}

Next we consider the symmetry class $T^2 = 1$ in 3+1D, corresponding to $\Omega^4_{pin-} = \Omega^4_{spin}(B\bZ_2,\xi)$ with $w_1(\xi) = A$, $w_2(\xi) = 0$, where $A$ is the $\bZ_2$ gauge field. It is known there are no phases in this symmetry class \cite{KTTW}.

The $E_2$ page of the AHSS has
\begin{equation}H^4(B\bZ_2,U(1)^\xi) = \bZ_2\end{equation}
\begin{equation}H^3(B\bZ_2,\Omega^1_{spin}) = \bZ_2\end{equation}
\begin{equation}H^2(B\bZ_2,\Omega^2_{spin}) = \bZ_2\end{equation}
\begin{equation}H^1(B\bZ_2,\Omega^3_{spin}) = \bZ_2,\end{equation}
so all these classes must be eaten either by incoming differentials from the 2+1D phases, or by outgoing differentials to the 4+1D phases. However, we are interested in $H^1(B\bZ_2,\Omega^3_{spin})$, which cannot be the target of a differential. Therefore, we are interested in differentials going to the 4+1D phases, for which the $E_2$ page has:
\begin{equation}H^5(B\bZ_2,U(1)^\xi) = 0\end{equation}
\begin{equation}H^4(B\bZ_2,\Omega^1_{spin}) = \bZ_2\end{equation}
\begin{equation}H^3(B\bZ_2,\Omega^2_{spin}) = \bZ_2\end{equation}
\begin{equation}H^2(B\bZ_2,\Omega^3_{spin}) = 0.\end{equation}

We see there are two possible targets for $\nu_3 = A$, either the generator of $H^3(B\bZ_2,\Omega^2_{spin})$ by $d_2$, or if $d_2$ vanishes then it must be the generator of $H^4(B\bZ_2,\Omega^1_{spin})$. On the other hand, the possible 3+1D Kitaev string phase in $H^2(B\bZ_2,\Omega^2_{spin})$ with $\nu_2 = A^2$ must also be eaten by a differential, and we can just compute using \eqref{e:2ndpostconstraintwisted}
\begin{equation}d_2 A^2 = Sq^2 A^2 + w_2(\xi) A^2 + w_1(\xi) Sq^1 A^2 = A^4 \neq 0.\end{equation}
Therefore, the possible target for $d_3\nu_3$ is already eaten by the $E_3$ page. Thus,
\begin{equation}d_2 \nu_3 = k_3 w_1(\xi)^2 \nu_3 = k_3 A^3 \neq 0,\end{equation}
which implies $k_3 = 1$.

\subsection{$\Omega^6_{pin+} = 0, k_1 = 1$}

Finally we consider the symmetry class $T^2 = (-1)^F$ in 5+1D. This corresponds to $\Omega^6_{pin+} = \Omega^6_{spin}(B\bZ_2,\xi)$ with $w_1(\xi) = A, w_2(\xi) = A^2$, where $A$ is the $\bZ_2$ gauge field. It is known that there are no such phases in this symmetry class and dimension \cite{KTTW}.

The $E_2$ page of the AHSS has
\begin{equation}H^6(B\bZ_2,U(1)^\xi) = \bZ_2\end{equation}
\begin{equation}H^5(B\bZ_2,\Omega^1_{spin}) = \bZ_2\end{equation}
\begin{equation}H^4(B\bZ_2,\Omega^2_{spin}) = \bZ_2\end{equation}
\begin{equation}H^3(B\bZ_2,\Omega^3_{spin}) = \bZ_2.\end{equation}
The $p+ip$ membrane phase $\nu_3 = A^3$ must either be eaten by an incoming differential or an outgoing differential. There are no possible incoming differentials so it has to be the later. Meanwhile, the next layer of the $E_2$ page has
\begin{equation}H^6(B\bZ_2,U(1)^\xi) = 0\end{equation}
\begin{equation}H^5(B\bZ_2,\Omega^1_{spin}) = \bZ_2\end{equation}
\begin{equation}H^4(B\bZ_2,\Omega^2_{spin}) = \bZ_2\end{equation}
\begin{equation}H^3(B\bZ_2,\Omega^3_{spin}) = 0,\end{equation}
so $\nu_3$ either has $d_2 \nu_3 \neq 0$ or $d_3 \nu_3 \neq 0$. However, we can just compute that $d_2 \nu_2 \neq 0$ as before using \eqref{e:2ndpostconstraintwisted}:
\begin{equation}d_2 A^4 = Sq^2 A^4 + A^6 = A^6 \neq 0.\end{equation}
Therefore, there is nowhere for $d_3 \nu_3$ to land, so $d_2 A^3 \neq 0$. Thus,
\begin{equation}d_2 A^3 = k_1 A^5 + A^5 + A^5 \neq 0,\end{equation}
from which we conclude $k_1 = 1$. This completes the derivation of the $d_2$ differential for the $p+ip$ part \eqref{e:3rdpostconstraint}.

\medskip

\bibliography{main}
\bibliographystyle{unsrt}

\end{document}